\documentclass[aps,amsfonts,amsmath,prd,preprint,nofootinbib]{revtex4}
\pdfoutput=1
\newcommand{\beq}{\begin{equation}}
\newcommand{\eeq}{\end{equation}}
\usepackage{graphicx}
\usepackage{natbib}
\usepackage[utf8]{inputenc}
\usepackage{subcaption}
\usepackage{MnSymbol}
\begin{document}

\title{Simulating cosmic string loop captured by a rotating black hole}
\author{Heling Deng}
\affiliation{Physics Department, Arizona State University, Tempe, AZ 85287, USA}
\affiliation{Department of Physics, Oregon State University, Corvallis, OR 97331, USA}
\author{Andrei Gruzinov}
\affiliation{Center for Cosmology and Particle Physics, Department of Physics, New York University, New York, NY 10001, USA}
\author{Yuri Levin}
\affiliation{Department of Physics, Columbia University, New York, NY 10027, USA}
\affiliation{Center for Computational Astrophysics, Flatiron Institute, New York, NY 10010, USA}
\affiliation{School of Physics and Astronomy, Monash Center for Astrophysics, Monash University, Clayton, VIC 3800, Australia}
\author{Alexander Vilenkin}
\affiliation{Institute of Cosmology, Department of Physics and Astronomy, Tufts University, Medford, MA 02155, USA}
\bigskip
\bigskip
\vspace{250px}
\begin{abstract}

We study the dynamics of a cosmic string loop captured by a rotating black hole, ignoring string reconnections. A loop is numerically evolved in Kerr spacetime, with the result that it turns into one or more growing
or contracting double-lines rotating around the black hole in the equatorial plane. {This is in good agreement with the approximate analytical treatment of the problem investigated by Xing et al., who studied the evolution of the auxiliary curve associated with the string loop. We} confirm that the auxiliary curve deformation can indeed describe the string motion in realistic physical scenarios to a reasonable accuracy, and can thus be used to further study other phenomena such as superradiance and reconnections of the captured loop.

\end{abstract}

\maketitle

\section{Introduction}

{Black holes and cosmic strings are fundamental geometric objects in theoretical physics. Black holes are non-trivial stationary solutions of the Einstein field equations; they can be formed by the collapse of massive dying stars and are ubiquitous in our universe, playing a significant role in a large number of astrophysical phenomena. The classical motion of a string can be described by the Nambu-Goto action. Although they have not been observed, cosmic strings can naturally arise in high energy physics as remnants of possible phase transitions in the early universe \cite{Kibble:1976sj,Vilenkin:2000jqa}. Oscillating string loops or a string network can generate potentially detectable phenomena, such as gravitational wave bursts \cite{Damour:2000wa,Damour:2004kw} and a stochastic gravitational wave background \cite{Vilenkin:1981bx,Damour:2004kw,Buchmuller:2020lbh,Ellis:2020ena,Blanco-Pillado:2021ygr,Hindmarsh:2022awe}. }

As the string network evolves and string loops float around in the universe, it is conceivable that they are scattered or even captured by black holes. {The interaction between these two fundamental objects has received relatively little attention in the literature. An infinitely long string or a circular string loop scattered or captured by a black hole were studied in refs. \cite{Lonsdale:1988xd,DeVilliers:1997nk,DeVilliers:1998nm,Frolov:1999pj,Snajdr:2002aa,Dubath:2006vs}. Refs. \cite{Frolov:1988zn,Frolov:1995vp,Igata:2018kry} found and investigated stationary string solutions in Kerr metric, demonstrating that the string can extract angular momentum from the rotating black hole.}

It has recently been argued in ref. \cite{Xing:2020ecz} that for a significant range of string tension, cosmic string loops are likely to be captured by supermassive black holes at the galactic centers.  It is also possible that supermassive black holes themselves were seeded by string loops: a gas cloud at redshifts $z=\mathcal{O}(100)$ could fall into a superconducting string loop and directly collapse into a massive black hole \cite{Cyr:2022urs}, which may capture the loop afterwards. In addition, if primordial black holes co-existed with cosmic strings in the early universe, it is expected that nearly all of them would end up with string loops attached to them {as a result of reconnections in the string network} \cite{Xing:2020ecz, Vilenkin:2018zol}.

\begin{figure}[!]
\includegraphics[scale=0.7]{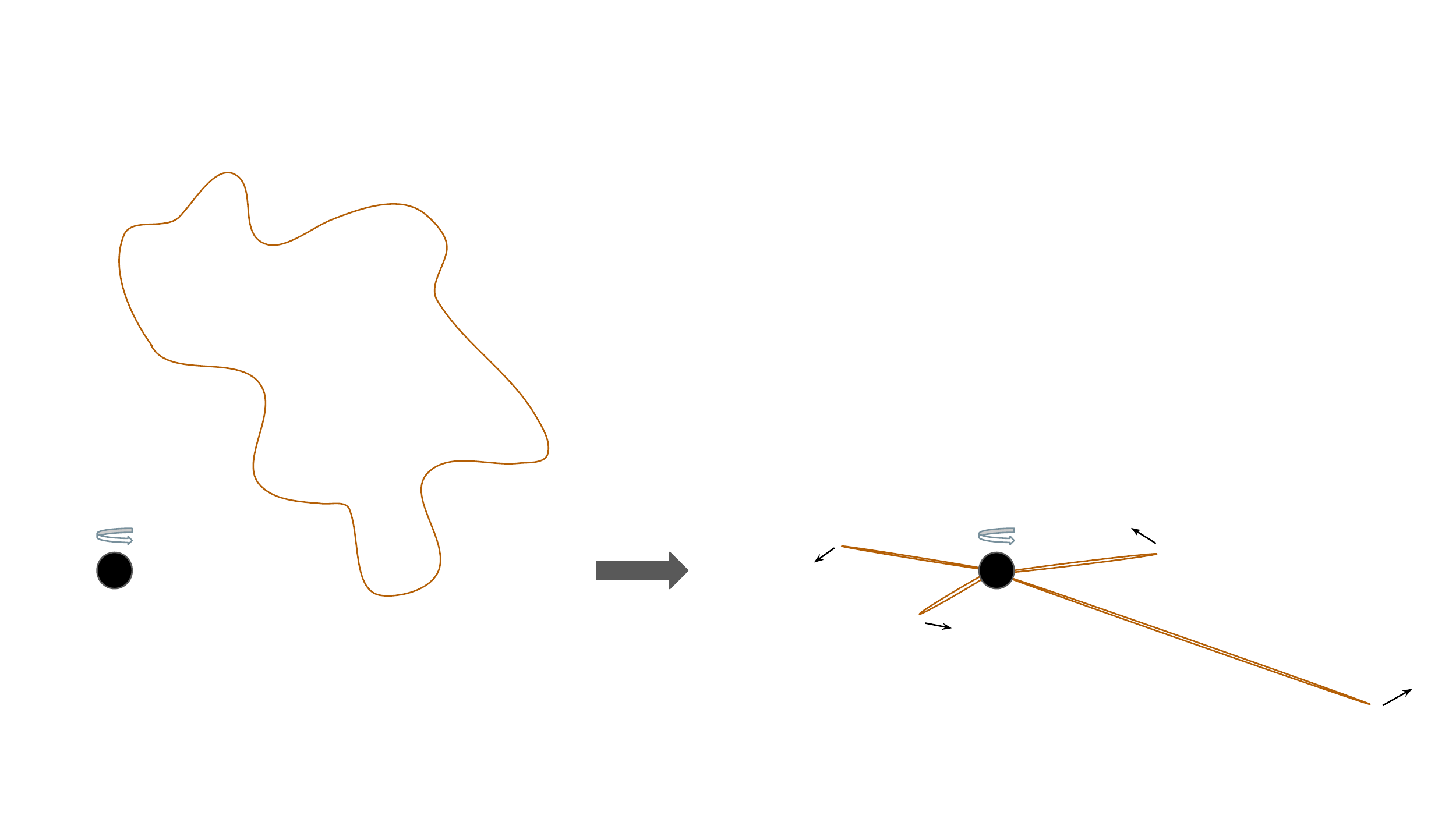}

\caption{\label{fig:string_introduction}Schematic pictures of the fate of a string loop near a rotating black hole. The loop may experience complex motion after being captured, but eventually it would turn into a simple structure: growing double-lines rotating in a direction close to that of the black hole spin. The tips of the double-lines move at a speed close to that of light.}
\end{figure}

Inspired by previous studies, {ref. \cite{Xing:2020ecz} investigated the evolution of a string loop captured by a rotating black hole significantly smaller in size than the loop. It was found that after several reconnections that took place on the loop oscillation timescale, the loop settled into a non-self-intersecting trajectory that evolved secularly due to the loop's interaction with the black hole. An approximate formalism was developed to study this process, where a mathematical tool called the auxiliary curve was introduced. Within this framework, it was predicted that the loop would either get fully swallowed by the black hole, or asymptotically turn into one or more lengthening double-lines} rotating close to the equatorial plane of the black hole spin, with one end attached to the black hole, and the tip moving at a speed close to that of light (fig. \ref{fig:string_introduction}). The lengths of the double-lines would keep growing until the black hole loses all of its rotational energy. These double-lines are possible sources of gravitational radiation.

In the present work, we apply numerical simulations to investigate the string dynamics in a realistic setting, and inspect the predictions made in ref. \cite{Xing:2020ecz}. We will show how the structure of double-lines as shown in fig. \ref{fig:string_introduction} arises, and how the estimates in ref. \cite{Xing:2020ecz} are indeed compatible with numerical experiments.

The rest of the paper is organized as follows. In sec. \ref{Auxiliary_curve} we review the concept of auxiliary curve introduced in ref. \cite{Xing:2020ecz}. In sec. \ref{sec:Simulating-string-black-hole} we detail our simulation setup and show the simulation results of string loops being captured by a rotating black hole. In sec. \ref{subsec:Superradiance} we briefly discuss the effect of superradiance on the captured string. Conclusions are summarized and further discussed in sec. \ref{discussion}. Supplementary materials are provided in the appendices. We set $c=G=1$ throughout the paper. 

\section{Auxiliary curve \label{Auxiliary_curve}}

The dynamics of a captured string loop can be found by simulating the string motion in black hole spacetime, the details of which will be shown and discussed in sec. \ref{sec:Simulating-string-black-hole}. In this section, we shall briefly review a mathematically convenient approach in estimating the string motion, with the help of the so-called auxiliary curve. This method was developed in ref. \cite{Xing:2020ecz}. As we will see later, though restricted by some assumptions, the approximation turns out to be consistent with simulations in general relativity.

\subsection{String reconstruction}

Let $M$ be the black hole mass. We first consider a simplified scenario, where a string with length
$L\gg M$ is moving in Minkowski spacetime, and two ends of the string
are attached to the black hole. A standard form of the string equation of motion in Minkowski spacetime is \cite{Vilenkin:2000jqa}
\begin{equation}
\ddot{\boldsymbol{X}}-\boldsymbol{X}^{\prime\prime}=0.\label{eq:eom in Minkowski}
\end{equation}
Here, $\boldsymbol{X}=\boldsymbol{X}(\sigma,t)$ is the position of
a string point in our three-dimensional space, where $\sigma$ is the
spatial coordinate marking points along the string, and $t$ is time.
We have defined $\dot{}\equiv\partial/\partial t$ and $^{\prime}\equiv\partial/\partial\sigma$.
This equation is obtained by specifying the so-called conformal gauge:
\begin{align}
\dot{\boldsymbol{X}}\cdot\boldsymbol{X}^{\prime} & =0,\label{eq:conformal gauge 1}\\
\dot{\boldsymbol{X}}^{2}+\boldsymbol{X}^{\prime2} & =1.\label{eq:conformal gauge 2}
\end{align}
In this gauge, the conserved total energy of the string is given by $E=\mu\int\text{d}\sigma=\mu L$, where $\mu$ is the constant string mass density (or the string tension). 

Of course, the string loop's energy is not conserved in the presence of the black hole. Part of the string would fall into the horizon, which decreases the string length; meanwhile, the string may extract energy from the black hole's rotation, which increases the string length. However, for $L\gg M$, the oscillation period of the string loop is much smaller than the time it takes for the string loop's energy to change significantly. 

Therefore, within an oscillation period, we can neglect
the black hole's gravitational effect, and the string loop simply evolves in
Minkowski spacetime with one point pinned at the black hole. 
From eq. (\ref{eq:eom in Minkowski}),
it can easily be shown that the motion of the free part is given by
\begin{equation}
\boldsymbol{X}(\sigma,t)=\frac{1}{2}\left[\boldsymbol{\mathsf{a}}(\sigma-t)-\boldsymbol{\mathsf{a}}(-\sigma-t)\right],\label{eq:auxiliary}
\end{equation}
where $\boldsymbol{\mathsf{a}}(\zeta)$ is a periodic function satisfying
$\boldsymbol{\mathsf{a}}(\zeta)=\boldsymbol{\mathsf{a}}(\zeta+2L)$
and $|\boldsymbol{\mathsf{a}}^{\prime}|=1$. Hence the string loop
oscillates around the black hole with period $2L$. The periodic function $\boldsymbol{\mathsf{a}}$ defines a closed curve in three dimensions called the auxiliary curve. Since
$|\boldsymbol{\mathsf{a}}^{\prime}|=1$, the parameter $\zeta$ denotes
the arc length along the curve. Once $\boldsymbol{\mathsf{a}}(\zeta)$
is known, we can tell the string configuration at any time $t$, with string length $L$.

\subsection{Torque from black hole}
The black hole can both absorb energy from and provide energy to the oscillating string loop, changing its length and shape. In order to take this into account, one can consider the total torque acting on the string, which effectively gives eq. (\ref{eq:eom in Minkowski}) two boundary conditions imposed respectively on the two string ends attached to the black hole. Since $L\gg M$, viewed from the vicinity of the black hole, a string stretches towards the horizon from a distance $\gg M$ on a nearly radial straight line, until it is curved by the black hole's rotational drag. Let $\boldsymbol{\Omega}_\text{string}=\boldsymbol{n}\times \dot{\boldsymbol{n}}$ be the angular velocity of the straight part of the string, where $\boldsymbol{n}$ is the unit vector pointing along the string in the direction away from the black hole. Suppose $\boldsymbol{\Omega}_\text{string}$ is much smaller than the angular velocity of the black hole $\boldsymbol{\Omega}_{\text{BH}}$, then the string can be regarded as quasi-stationary. The torque it receives from the black hole's rotational drag is $\boldsymbol{Q}_{\text{RD}}=-(\mu/M)(\boldsymbol{J}\times\boldsymbol{n})\times\boldsymbol{n}$, where $\boldsymbol{J} $ is the angular momentum of the black hole. This comes from the analytic solution of a stationary string in Kerr spacetime, first found in ref. \cite{Frolov:1988zn}.

In addition, the slowly rotating string experiences another torque that comes from the so-called ``horizon friction'' \cite{Xing:2020ecz}. For a small black hole spin ($M{\Omega}_\text{BH}\ll 1$), we expect that this torque $\boldsymbol{Q}_\text{HF}\propto \boldsymbol{\Omega}_\text{string}$.\footnote{\label{footnote} To the lowest order in $\boldsymbol{\Omega}_\text{string}$, $\boldsymbol{Q}_\text{HF}$ may also contain other terms such as $\sim (\boldsymbol{\Omega}_\text{string} \cdot \boldsymbol{\Omega}_\text{BH})\boldsymbol{\Omega}_\text{BH}$ and $\sim \boldsymbol{\Omega}_\text{string} \times \boldsymbol{\Omega}_\text{BH}$. These terms are subordinate compared to $\sim \boldsymbol{\Omega}_\text{string}$ when $M{\Omega}_\text{BH}\ll 1$.} The prefactor can be determined by requiring that the straight string co-rotates with the black hole when the total torque acting on it vanishes, i.e., $\boldsymbol{\Omega}_\text{string}=\boldsymbol{\Omega}_{\text{BH}}\approx \boldsymbol{J}/4M^3$ when $\boldsymbol{Q}=\boldsymbol{Q}_{\text{RD}}+\boldsymbol{Q}_\text{HF}=0$. This gives $\boldsymbol{Q}_\text{HF}=4\mu M^2\boldsymbol{\Omega}_\text{string}$.

Therefore, the total torque exerting on a string slowly rotating around a slowly rotating black hole can be estimated as
\begin{equation}
\boldsymbol{Q}=\frac{\mu}{M}\left(4M^{3}\dot{\boldsymbol{n}}-\boldsymbol{J}\times\boldsymbol{n}\right)\times\boldsymbol{n}.\label{eq:Q}
\end{equation}

\subsection{Auxiliary curve equation \label{subsec:auxiliary-curve-equation}}

The torque $\boldsymbol{Q}$ applied to the string changes its energy and angular momentum. As proposed in ref.  \cite{Xing:2020ecz}, it also changes the shape of the auxiliary curve $\boldsymbol{\mathsf{a}}(\zeta)$  –  the auxiliary curve moves with velocity $\boldsymbol{\mathsf{v}}(\zeta)$ given by\footnote{Note that there is a correction of sign compared to the corresponding equation in ref. \cite{Xing:2020ecz}.} \cite{Xing:2020ecz}:
\begin{equation}
\boldsymbol{\mathsf{v}}=\frac{4}{ML_{\mathsf{a}}}\left(4M^{3}\boldsymbol{\mathsf{a}}^{\prime\prime}+\boldsymbol{J}\times\boldsymbol{\mathsf{a}}^{\prime}\right),\label{eq:eom of auxiliary}
\end{equation}
where  $L_{\mathsf{a}}$ is the length of the auxiliary curve and the derivatives are understood to be evaluated with respect to $\zeta$.

Therefore, instead of solving eq. (\ref{eq:eom in Minkowski})
with boundary conditions, we can evolve this simple equation and find $\boldsymbol{\mathsf{a}}(\zeta)$ at any time, then use
eq. (\ref{eq:auxiliary}) to reconstruct the corresponding string configurations within a timescale $\sim L_\mathsf{a}$, during which the string has length $\sim L_{\mathsf{a}}/2$.\footnote{When the evolution of the auxiliary curve is taken into account, there is no longer a strict definition of oscillation period for the string. But since the curve is expected to stay almost unchanged within time $\sim L_\mathsf{a}$, it can be used to find the approximate string configurations within this timescale.} The change of $L_{\mathsf{a}}(t)$
is given by
\begin{equation}
\dot{L}_{\mathsf{a}}=-\oint\boldsymbol{\mathsf{v}}\cdot\boldsymbol{\mathsf{a}}^{\prime\prime}\text{d}\zeta.
\end{equation}
This can be used as an accuracy check when solving eq. (\ref{eq:eom of auxiliary})
numerically.

\subsection{Predictions from auxiliary curve \label{predictions}}
If the curve is initially smooth enough, its evolution is particularly simple. Several typical scenarios are shown in appendix A. Roughly speaking, the term containing $\boldsymbol{\mathsf{a}}^{\prime\prime}$ in eq. (\ref{eq:eom of auxiliary}) (the horizon friction term) tends to shrink the curve into a circle, whereas the term containing $\boldsymbol{J}$ (the rotational drag term) tends to turn the curve into an ever-expanding circle lying in the black hole's equatorial plane. By eq. (\ref{eq:auxiliary}), a circular auxiliary curve of length
$L_{\mathsf{a}}=2L$ corresponds to a string loop that extends radially
from the black hole to radius $L/\pi$ and then traces the same radial
line back to the black hole. This double-line rotates around
the black hole with angular velocity $\pi/L$, and its tip
moves at the speed of light.

Therefore, no matter which term dominates in the auxiliary curve equation, the corresponding string loop would eventually turn into a shortening/lengthening, rotating double-line with one end attached to the black hole. If the rotational drag term dominates, the double-line would eventually be rotating in the equatorial plane.\footnote{{In general, we would expect that more than one string segments are captured, and so a string loop would turn into a number of growing double-lines.}}

It should be stressed again that the derivation of the horizon friction term ($\sim \boldsymbol{\mathsf{a}}^{\prime\prime}$) in the auxiliary curve equation (or $\boldsymbol{Q}_\text{HF}$ in $\boldsymbol{Q}$) assumes a small black hole spin. We would now argue that this term becomes insignificant in the case of large black hole spin. To the order of magnitude, the first term inside the bracket in eq. (\ref{eq:eom of auxiliary}) is $\sim (M/L)M^2$, while the second term is $\sim M^2$ for a large spin. As long as $L \gg M$, which is the case in realistic physical scenarios, the rotational drag term dominates, and so whether or not the horizon friction is accurately accounted for becomes irrelevant. Therefore, the auxiliary curve equation is expected to hold in general. 

On the other hand, the derivation of this equation involves a number of approximations and assumptions which are heuristically plausible but do not have a rigorous justification.  This includes the approximation of the loop as moving in Minkowski space with a fixed point at black hole location and the assumptions we made about the torque due to horizon friction (see footnote \ref{footnote}). 

Does the auxiliary curve equation give only a qualitative picture of string motion, or a quantitative description? To figure this out is a goal of this work. In the next section, we will directly evolve string loops in black hole spacetime, inspecting the aforementioned  predictions from auxiliary curve deformation.

\section{Simulating string-black hole interaction\label{sec:Simulating-string-black-hole}}

In this section, we show how we applied numerical simulations to study
a string loop interacting with a rotating black hole. We will see 
how a captured loop turns into lengthening/shortening rotating double-lines as it oscillates around the black hole. The results will be compared with those from the method of auxiliary curve.

\subsection{Black hole spacetime as background}

We simulated the evolution of a Nambu-Goto string in
a fixed black hole background. This is plausible as long as the gravitational
effect from the string is negligible, i.e., the string has a mass
much smaller than the black hole mass. 

The line element of Kerr spacetime in the standard Boyer-Lindquist coordinates is given by
\begin{equation}
\text{d}s^{2}=-\left(1-\frac{2Mr}{\rho^{2}}\right)^{2}\text{d}t^{2}-\frac{4Mar\sin^{2}\theta}{\rho^{2}}\text{d}t\text{d}\phi+\frac{\Sigma}{\rho^{2}}\sin^{2}\theta\text{d}\phi^{2}+\frac{\rho^{2}}{\Delta}\text{d}r^{2}+\rho^{2}\text{d}\theta^{2},
\end{equation}
where 
\begin{align}
\rho^{2} & =r^{2}+a^{2}\cos^{2}\theta,\\
\Delta & =r^{2}-2Mr+a^{2},\\
\Sigma & =\left(r^{2}+a^{2}\right)^{2}-a^{2}\Delta\sin^{2}\theta.
\end{align}
Here $a$ is known as the Kerr parameter. The angular momentum of
the black hole is $\boldsymbol{J}=aM\hat{\boldsymbol{z}}$, with $|a|<M$.
The outer horizon of the black hole is located at
\begin{equation}
r=M+\sqrt{M^{2}-a^{2}}.
\end{equation}
It would be more convenient to simulate the string using Cartesian
coordinates. The Kerr-Schild metric is a possible option, but we found
it more convenient to use the so-called quasi-isotropic metric (see, e.g., ref. \cite{Brandt:1996si}), where
the string gets ``frozen'' as it approaches the outer horizon and
hence avoids any singularities. To this end, we first perform the
following transformation, replacing
$r$ by $\bar{r}$ in Boyer-Lindquist coordinates:
\begin{equation}
r=\bar{r}\left(1+\frac{M+a}{2\bar{r}}\right)\left(1+\frac{M-a}{2\bar{r}}\right).
\end{equation}
The line element is then of a quasi-isotropic form
\begin{equation}
\begin{split}
\text{d}s^{2} & =-\left(1-\frac{2Mr}{\rho^{2}}\right)\text{d}t^{2}+\left(\frac{\rho}{\bar{r}}\right)^{2}\left[\text{d}\bar{r}^{2}+\bar{r}^{2}\left(\text{d}\theta^{2}+\sin^{2}\theta\text{d}\phi^{2}\right)\right]\\
 & -\frac{4Mar\sin^{2}\theta}{\rho^{2}}\text{d}t\text{d}\phi+\left(1+\frac{2Mr}{\rho^{2}}\right)a^{2}\sin^{4}\theta\text{d}\phi^{2}.
\end{split}
\end{equation}
The outer horizon is given by
\begin{equation}
\bar{r}=\frac{\sqrt{M^{2}-a^{2}}}{2}.\label{eq:horizon}
\end{equation}
In order to express the line element in Cartesian coordinates, let
\begin{align}
x & =\bar{r}\sin\theta\cos\phi,\\
y & =\bar{r}\sin\theta\sin\phi,\\
z & =\bar{r}\cos\theta.
\end{align}
Then we get
\begin{equation} \label{eq:iso}
\begin{split}
\text{d}s^{2} & =-\left(1-\frac{2Mr}{\rho^{2}}\right)\text{d}t^{2}+\left(\frac{\rho}{\bar{r}}\right)^{2}\left(\text{d}x^{2}+\text{d}y^{2}+\text{d}z^{2}\right)\\
 &  +\frac{4Mr}{\rho^{2}}\frac{a}{\bar{r}^{2}}\left(y\text{d}x-x\text{d}y\right)\text{d}t+\left(1+\frac{2Mr}{\rho^{2}}\right)\left(\frac{a}{\bar{r}^{2}}\right)^{2}\left(y\text{d}x-x\text{d}y\right)^{2}.
\end{split}
\end{equation}
This is the metric we use to evolve the string motion.
If $a=0$, it reduces to the line element of a Schwarzschild black
hole expressed in the well-known isotropic coordinates,
\begin{equation}
\text{d}s^{2}=-\frac{\left(1-\frac{M}{2\bar{r}}\right)^{2}}{\left(1+\frac{M}{2\bar{r}}\right)^{2}}\text{d}t^{2}+\left(1+\frac{M}{2\bar{r}}\right)^{2}\left(\text{d}x^{2}+\text{d}y^{2}+\text{d}z^{2}\right),
\end{equation}
which can be used for simulations in the case of non-rotating black
hole. 

\subsection{String equation of motion}

Having specified the spacetime background $g_{\alpha\beta}$, we need
to write down the string equation of evolution, which should be the
generalization of eq. (\ref{eq:eom in Minkowski}). A one-dimensional
string can be represented by a two-dimensional surface in spacetime,
\begin{equation}
x^{\nu}=x^{\nu}(\zeta^{i}),
\end{equation}
where $\nu=0,1,2,3$ and $i=0,1.$ This is called the string worldsheet.
The coordinates $\zeta^{i}$ are arbitrary parameters: $\zeta^{0}$
is chosen to be timelike, while $\zeta^{1}$ is spacelike. The two-dimensional
worldsheet metric is given by
\begin{equation}
\gamma_{ij}=g_{\alpha\beta}\frac{\partial x^{\alpha}}{\partial\zeta^{i}}\frac{\partial x^{\beta}}{\partial\zeta^{j}}.
\end{equation}
The contravariant metric tensor $\gamma^{ij}$ can be defined by $\gamma^{ij}\gamma_{jk}=\delta_{k}^{i}$. 
{The string equation of motion can
then be obtained from the Nambu-Goto action, and is given by \cite{Vilenkin:2000jqa}}
\begin{equation}
\frac{1}{\sqrt{-\gamma}}\frac{\partial}{\partial\zeta^{i}}\left(\sqrt{-\gamma}\gamma^{ij}\frac{\partial x^{\nu}}{\partial\zeta^{j}}\right)+\Gamma_{\alpha\beta}^{\nu}\gamma^{ij}\frac{\partial x^{\alpha}}{\partial\zeta^{i}}\frac{\partial x^{\beta}}{\partial\zeta^{j}}=0,\label{eq:stringeom}
\end{equation}
where $\gamma \equiv \det(\gamma_{ij})$ and $\Gamma_{\alpha\beta}^{\nu}$ is the Christoffel symbol associated
with $g_{\alpha\beta}$.

Now, it is convenient to set $\zeta^{0}=t$, which is simply the time
for the four-dimensional spacetime. Let $\sigma\equiv\zeta^{1},$
then $\boldsymbol{x}(\sigma,t)$ is the position of a string point
$\sigma$ at time $t$. Imposing a gauge condition $\dot{x}\cdot x^{\prime}=0$,
the string worldsheet metric $\gamma_{\alpha\beta}$ can be found
as $\gamma_{00}=\dot{x}^{2}$, $\gamma_{11}=x^{\prime2}$ and $\gamma_{01}=0$.
The string equation of motion (\ref{eq:stringeom}) then becomes  \cite{Vilenkin:2000jqa}
\begin{equation}
\partial_{t}\left(F^{-1}\dot{x}^{\nu}\right)-\left(Fx^{\nu\prime}\right)^{\prime}+\Gamma_{\alpha\beta}^{\nu}\left(F^{-1}\dot{x}^{\alpha}\dot{x}^{\beta}-Fx^{\alpha\prime}x^{\beta\prime}\right)=0,\label{eq:eom2}
\end{equation}
where we have defined
\begin{equation}
F\equiv\sqrt{-\frac{\dot{x}^{2}}{x^{\prime2}}}.
\end{equation}
We solve equations (\ref{eq:eom2}) to find $\boldsymbol{x}=(x,y,z)$
as functions of $t$ and $\sigma$. In addition, although $F(\sigma)$
can in principle be given by $\boldsymbol{x}(\sigma)$ and $\dot{\boldsymbol{x}}(\sigma)$
from its definition, we found it computationally more  convenient to
evolve $F$ as well. The equation for $F$ is simply the 0-component
of eq. (\ref{eq:eom2}).

\subsection{Simulation setup}

{In our simulations, we start with a string loop that is not yet captured by the black hole, and then model both the process of the capture and the subsequent evolution of the loop. We use the standard finite-difference method to numerically solve
eq. (\ref{eq:eom2})}, where $\boldsymbol{x}$ and
$F$ are ``fields'' evolving on the one-dimensional string. Since
we are dealing with string loops, we need to impose periodic boundary
conditions. The initial conditions are less trivial: we need to find
$\boldsymbol{x}(\sigma)$, $\dot{\boldsymbol{x}}(\sigma)$, $F(\sigma)$
and $\dot{F}(\sigma)$ at the initial time that satisfy
\begin{equation}
\dot{x}\cdot x^{\prime}=0,\label{eq:gauge}
\end{equation}
\begin{equation}
\dot{x}^{2}+F^{2}x^{\prime2}=0,\label{eq:def of F}
\end{equation}
where the first equation is our imposed gauge condition, and the second
one is from the definition of $F$. Note that these two conditions
look very similar to the conformal gauge for strings living in Minkowski
spacetime (eqs. (\ref{eq:conformal gauge 1}) and (\ref{eq:conformal gauge 2})),
where a number of analytic solutions $\boldsymbol{X}(\sigma,t)$ have
been found in the literature (e.g., refs. \cite{Kibble:1982cb,Burden:1985md}). It would be convenient if we
could modify these known solutions and turn
them into the initial conditions needed in our simulations. By doing so, we can place the loop near the black hole at the
initial time, and the runtime of simulations can thus be reduced significantly.\footnote{In some previous works (e.g., ref. \cite{Snajdr:2002aa}) the initial string was placed at a large distance from the black hole such that the string motion can be studied in the weak-field limit.} While the way to obtain $\boldsymbol{x}$ and $\dot{\boldsymbol{x}}$
from $\boldsymbol{X}(\sigma,t)$ is not unique, we found the following
transformations should work well for our purposes:
\begin{align}
\boldsymbol{x} & =\boldsymbol{X},\label{eq:from X to x}\\
\frac{\dot{\boldsymbol{x}}}{F} & =\dot{\boldsymbol{X}}+B\boldsymbol{X}^{\prime},\label{eq:from Xdot to xdot}
\end{align}
where $B$ is a function of $\boldsymbol{X},\dot{\boldsymbol{X}}$
and $\boldsymbol{X}^{\prime}$, and can be found by solving eqs. (\ref{eq:gauge})
and (\ref{eq:def of F}). The expressions of $B$ and $F$ are rather complicated due to the complicated Kerr metric. It is shown with more details in appendix B how they are obtained. 

An important fact about our formalism is that eq. (\ref{eq:gauge})
and $\zeta^{0}=t$ do not fix the gauge completely. We still have
the freedom to perform transformations $\sigma\to\tilde{\sigma}(\sigma)$.
This can be seen by observing that terms with $Fx^{\alpha\prime}$
in eq. (\ref{eq:eom2}) stay invariant under the transformation of
$\sigma$. In simulations, this means we can freely add 
points in our grid (with grid spacing unchanged\footnote{This means we need not perform adaptive refinement in either grid spacing or time step.}) as long as $Fx^{\alpha\prime}$
is invariant at each string point. This is especially useful
when dealing with a lengthening string: one can simply add more grid
points to the segment being stretched (near the black hole horizon), such that a sufficient
numerical resolution is guaranteed.\footnote{In some earlier works studying the string-black hole interaction (e.g., refs. \cite{DeVilliers:1998nm,Frolov:1999pj,Snajdr:2002aa}), the string was described in the conformal gauge, i.e., $\dot{x}\cdot x^\prime = 0$ and $\dot{x}^2+x^{\prime 2}=0$, where $\dot{} \equiv \partial/\partial \zeta^0$ and $^\prime \equiv \partial/\partial \zeta^1$. By adopting these two conditions, the residual gauge freedom no longer allows $\zeta^0=t$. In our formalism, we abandon the second condition and replace it by $\zeta^0=t$. This not only provides a more physical description of the string, but also allows us to freely transform the spacelike coordinate $\sigma$, which is extremely helpful when adding grid points to a lengthening string segment.}

In order to evaluate the accuracy of our numerical results, we can check
the ``invariant length'' of the string loop, which is basically
a check of energy conservation. To begin with, the conservation of
the energy-momentum tensor $T^{\alpha\beta}$ gives
\begin{equation}
\partial_{\nu}\left(T_{\alpha}^{\nu}\sqrt{-g}\right)-\frac{1}{2}\sqrt{-g}T^{\nu\rho}\partial_{\alpha}g_{\nu\rho}=0.
\end{equation}
In stationary spacetime, $\partial_{0}g_{\nu\rho}=0$, and it follows
that for a localized matter distribution, the total energy
\begin{equation}
E\equiv\int T_{0}^{0}\sqrt{-g}\text{d}^{3}x=\text{const}.
\end{equation}
Specific expressions for $T^{\alpha\beta}$ of the string can be found as \cite{Vilenkin:2000jqa}
\begin{equation}
T^{\alpha\beta}\sqrt{-g}=\mu\int\text{d}^{2}\zeta\sqrt{-\gamma}\gamma^{ij}x_{,i}^{\alpha}x_{,j}^{\beta}\delta^{(4)}\left(x^{\nu}-x^{\nu}(\zeta^{i})\right)\label{eq:EM tensor}
\end{equation}
Imposing our gauge condition (\ref{eq:gauge}) and $\zeta^{0}=t$,
we finally get
\begin{equation}
E=\mu\int\frac{\dot{x}_{0}}{F}\text{d}\sigma,\label{eq:invariant length}
\end{equation}
where $\dot{x}_{0}=g_{0\alpha}\dot{x}^{\alpha}$. Since $E=\text{const}$,
we define $L_{i}\equiv E/\mu$ as the invariant length of the string
loop. In static spacetime (where $g_{0\alpha}=0$ except for $g_{00}$), $\dot{x}_{0}=g_{00}$, and, from the 0-component
of the string equation of motion (\ref{eq:eom2}), $\partial_{t}(g_{00}/F)=0$. Therefore, we have $L_{i}\propto\int\text{d}\sigma$
up to a constant factor, which is consistent with the commonly used definition in Minkowski  spacetime.

At a first glance, $E=\text{const}$ seems incompatible with the concept
of ``string-lengthening/shortening'' discussed previously. When we say ``string-shortening''
(which is due to horizon friction), we imagine that part of the string
falls into the horizon as it rotates around the black hole. This is
no longer the case when we use the (quasi-)isotropic coordinates,
because it takes forever for the string to get to the horizon.
Instead, the string would ``wrap around'' the black hole, much like
a strand of yarn wrapping around a yarn ball. If we include the part
``frozen'' on the horizon, the total energy of the string loop is
indeed conserved. ``String-shortening'' only refers to the part
``away'' from the black hole horizon. The definition of ``away''
is rather arbitrary, because the whole string is technically always
outside the horizon. When discussing the change of the string length, we only
take into account the string segment whose distance from the black
hole center is larger than, say, 1.1 times the horizon. The exact
length scale here is not very important as long as it is much smaller than the string length. Similarly, for a string loop experiencing
``string-lengthening'', we neglect the part that is already attached
to the black hole's outer horizon\footnote{Apart from being ``frozen'' on the outer horizon, this part of string is also co-rotating
with the black hole.}. In fact, this ``frozen'' part has negative energy, compensating
the growing part outside the black hole. The total energy
defined by eq. (\ref{eq:invariant length}) in conserved.

A related numerical issue occurs when a string segment is very
close to the outer horizon. Theoretically, the string gets asymptotically
close to the horizon and never reaches it. In simulations, however,
due to inevitable errors from the grid setup, a grid point on the
string may get inside the horizon, leading to numerical
instabilities. To avoid this, we stop evolving a grid point when it is ``sufficiently''
close to the horizon, which, again, can be defined more or less arbitrarily. In our experiments,
we manually force the smallest distance from a grid point to the black
hole center to be $1+10^{-5}$ times the outer horizon. A string segment
is said to be ``captured'' once it reaches this minimum radius.

\subsection{Two simulated examples: initial conditions}

In what follows, we will consider two examples from our simulations, one with a relatively large black hole spin and the other with a small spin. The initial string configurations for the two examples are shown in blue in fig. \ref{fig:string_before_capture}.

To set up the initial conditions, let us first consider a circular string loop in Minkowski spacetime:
\begin{align}
\boldsymbol{X}(\sigma,t) & =\frac{L_{i}}{4\pi}\left(\sin\sigma_- +\sin\sigma_+,-\cos\sigma_- -\cos\sigma_+,0\right),\label{eq:Minkowski loop}
\end{align}
where $\sigma_+ \equiv (2\pi/L_i)(\sigma+t)$, $\sigma_- \equiv (2\pi/L_i)(\sigma-t)$, and $L_i$ is a constant parameter. This is a solution of eqs. (\ref{eq:eom in Minkowski}-\ref{eq:conformal gauge 2}). It describes
an oscillating circular loop of period $L_i$ in the $z=0$ plane that collapses to a point and then re-expands
into a circle. The loop has an invariant
length $L_{i}$. At time $t=0$,
the circle has a maximum radius $L_{i}/2\pi$, and the string velocity
$\dot{\boldsymbol{X}}$ vanishes everywhere.

In our simulations, a black hole of mass $M$ is centered at $(0,0,0)$, and its angular momentum is set to be parallel to the $z$-axis. The invariant length of the string loop in eq. (\ref{eq:Minkowski loop}) is set to be $L_{i}=300M$. 

We consider the following two examples. (A) The black hole's Kerr parameter is set to be $a=-0.8M$, so the spin is relatively large. The center of the initial loop is shifted to $\frac{L_i}{4\pi}(1,0.5,0.1)$. The circle lies in a plane close to the plane of $z=0$. (B) The Kerr parameter is set to be $a=-0.1M$, hence a relatively small black hole spin. The center of the initial loop is shifted to $\frac{L_i}{4\pi}(1,0.5,0)$; then the loop is tilted by an angle $\pi/4$ about the diameter parallel to the $y$-axis such that its initial angular momentum does not align with the black hole spin. These two examples will be referred to as example A and example B, respectively.

After the Minkowski solutions $\boldsymbol{X}$ and $\dot{\boldsymbol{X}}$ are specified, we transform them into the initial
$\boldsymbol{x}$ and $\dot{\boldsymbol{x}}$ for Kerr spacetime according
to eqs. (\ref{eq:from X to x}) and (\ref{eq:from Xdot to xdot}). Strictly speaking, the invariant length is no longer (exactly) $L_{i}$, but the difference is insignificant when $L_i \gg M$.

\subsection{Results}

\subsubsection{String-lengthening/shortening}

\begin{figure}[!]
\centering

\subfloat[Example A]{\includegraphics[scale=0.35]{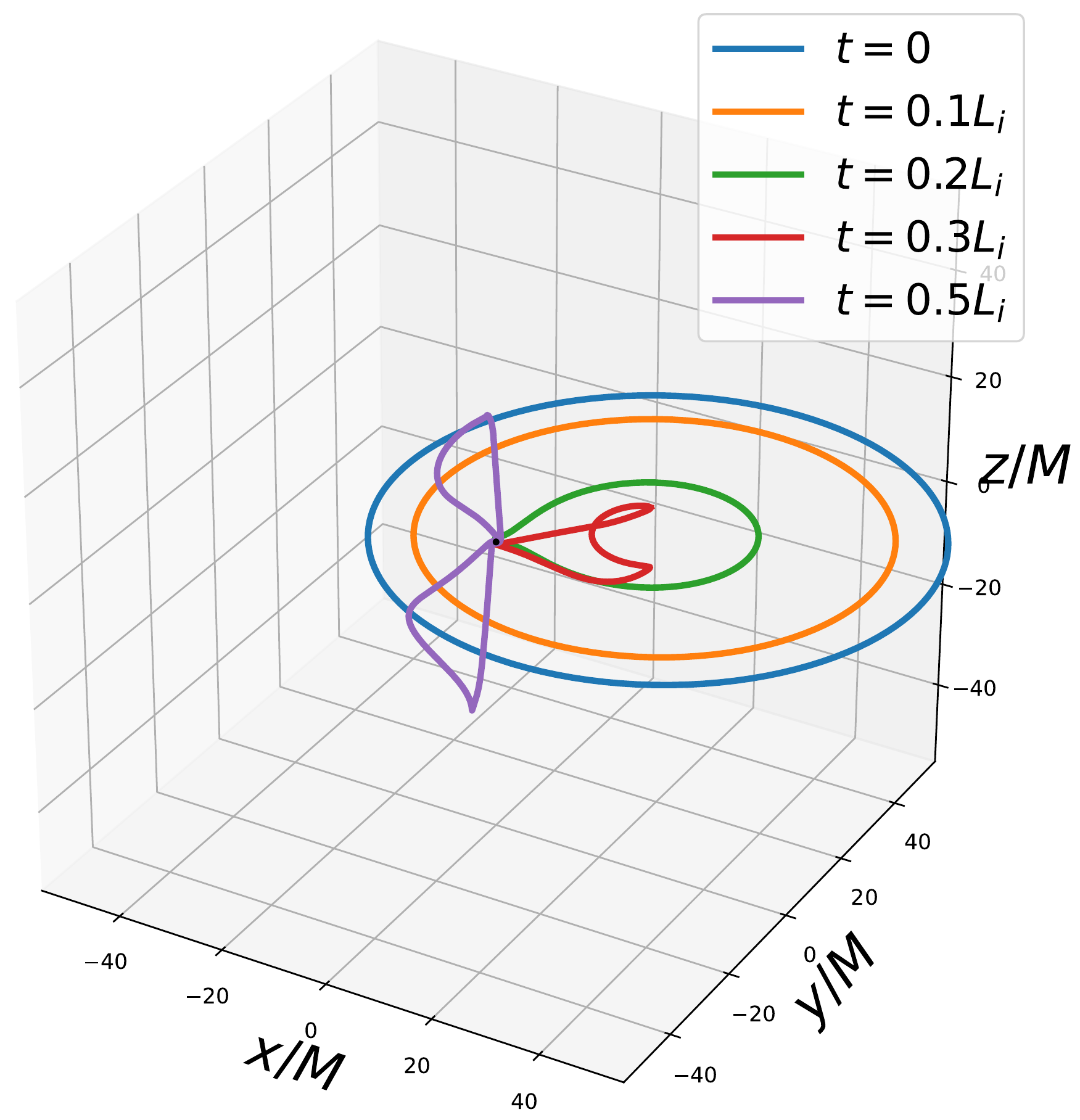}}\hspace{1.2cm}
\subfloat[Example B]{\includegraphics[scale=0.35]{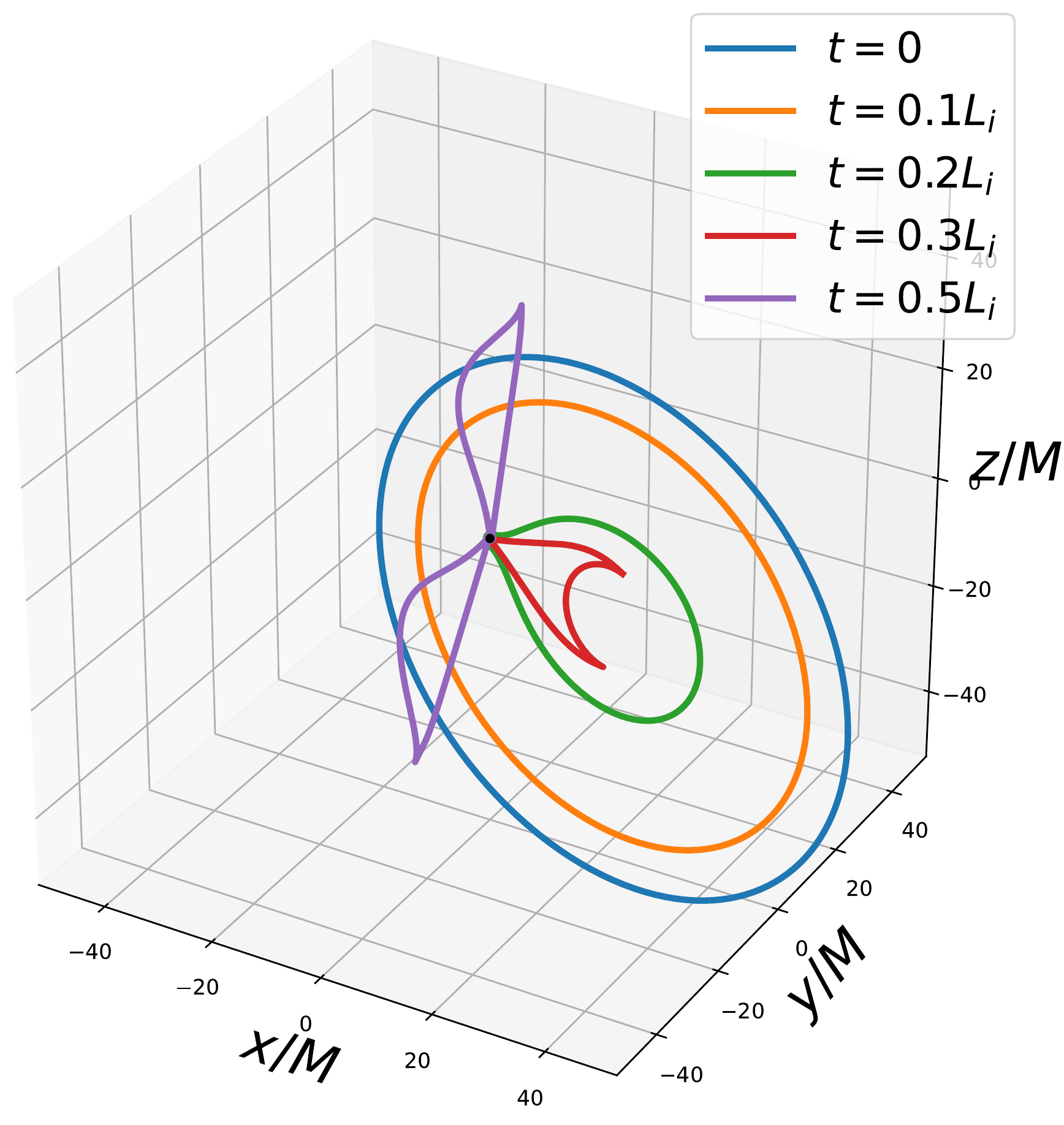}}
\protect\caption{\label{fig:string_before_capture}Several snapshots of string configuration
in the early stage of the interaction. The black hole is centered
at $(0,0,0)$ and is depicted as a sphere with radius given by eq. (\ref{eq:horizon}), which is the outer horizon (and is much smaller than the scale of the box in the figures so it appears to be a point at the center).
In both examples, the initial string loop is a circle with perimeter $L_{i}\approx300M$.
The loop is attracted towards the black hole, and
a string segment is captured at time $t\approx0.3L_{i}$ (red). Then
at $t\approx0.5L_{i}$, another segment is captured, and two loops
are formed (purple). One of the them (the one in region $y<0$) then
starts rotating clockwise almost in the plane of the initial circle, while the other rotates counterclockwise.}
\end{figure}

Several snapshots of the string configuration in the early stage of the two examples are shown in fig. \ref{fig:string_before_capture}. After the simulation sets off, the string loop starts to shrink (blue $\to$ orange $\to$ green). In
the meantime, part of the string is attracted towards the black hole. A string segment is captured at
time $t\approx0.3L_{i}$ (red); then another segment is captured at $t\approx0.5L_{i}$ (purple).
As a result, two sub-loops are formed, and their initial
lengths are both $\sim L_{i}/2$. The two loops then independently rotate almost in the plane of the initial circle, with one loop rotating clockwise and
the other counterclockwise.

\begin{figure}[!]
\subfloat[$t\approx0.9L_{i}$]{\includegraphics[scale=0.5]{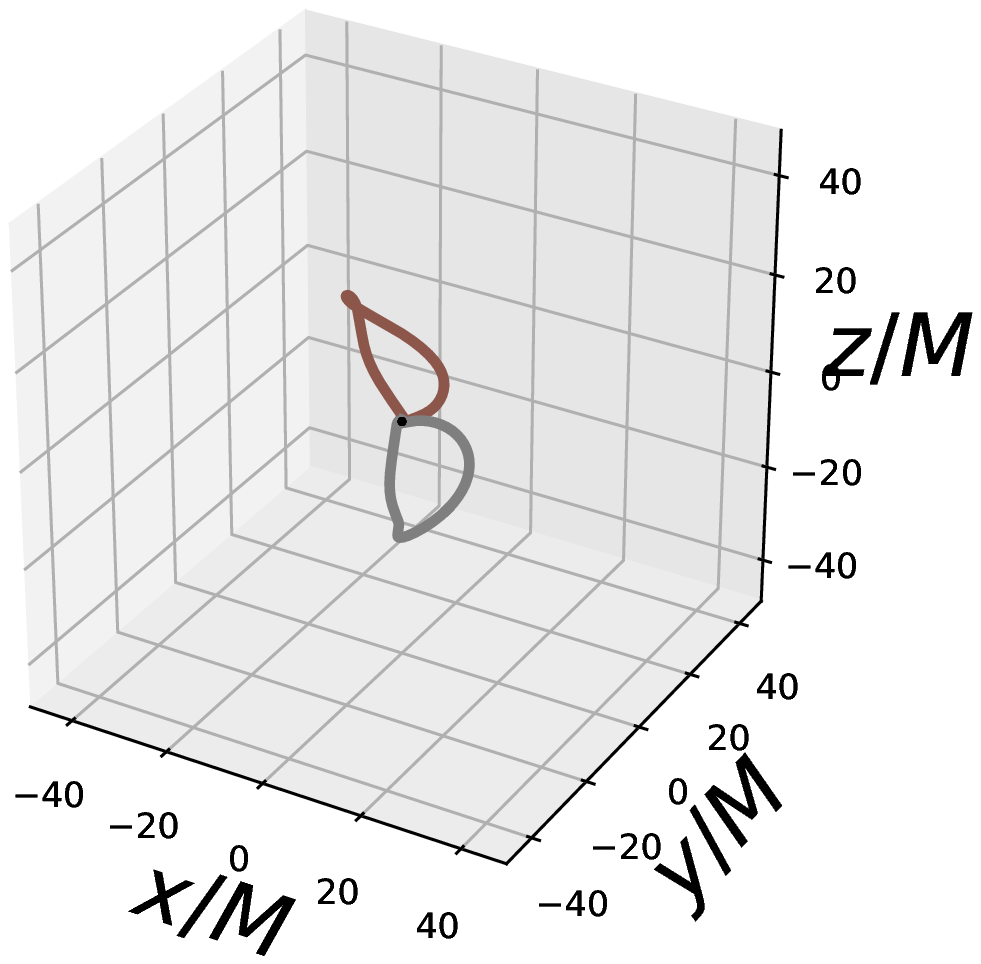}
}\hspace{-1.09cm}\subfloat[$t\approx2.7L_{i}$]{\includegraphics[scale=0.5]{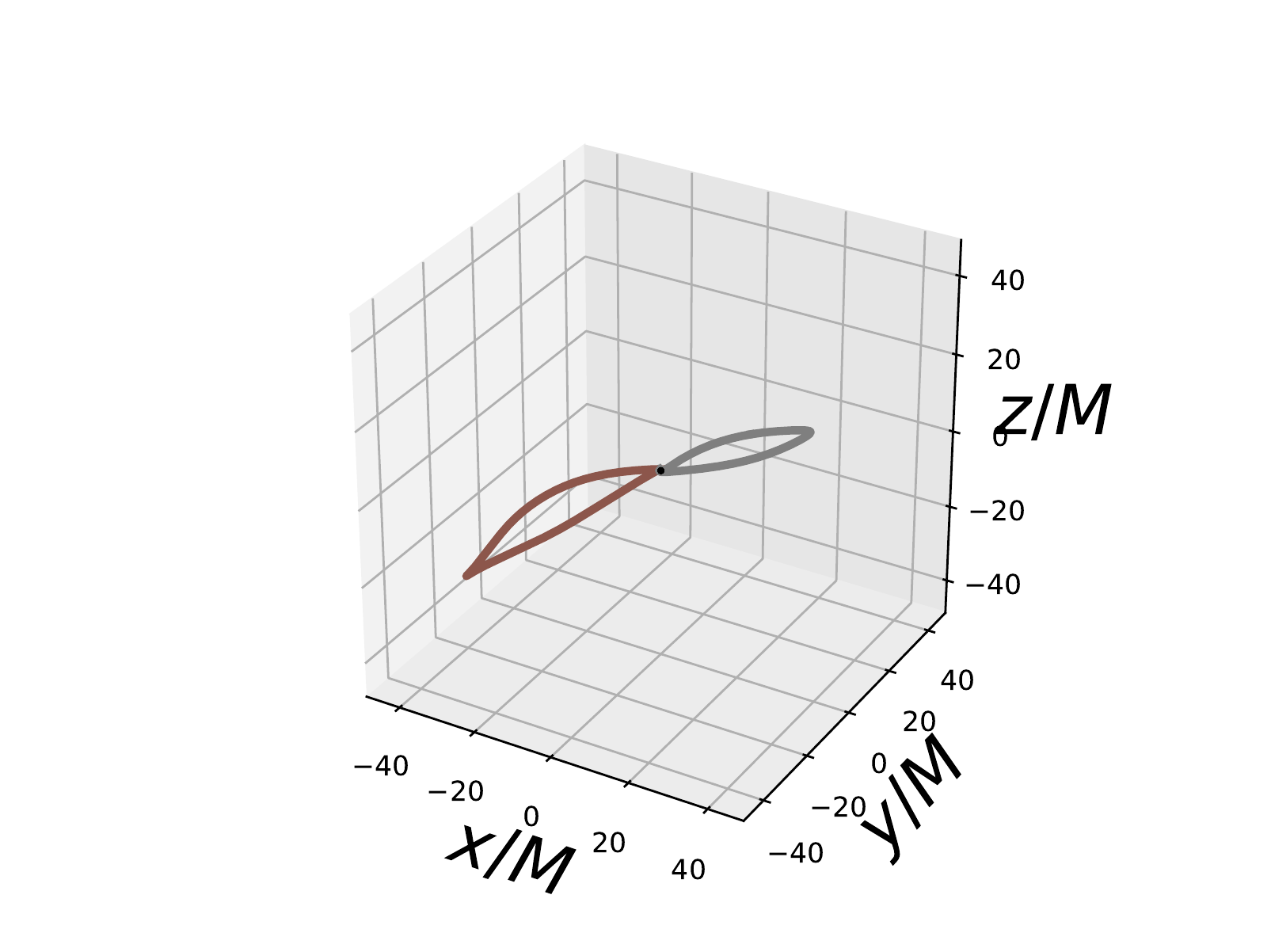}}

\subfloat[$t\approx4L_{i}$]{\includegraphics[scale=0.5]{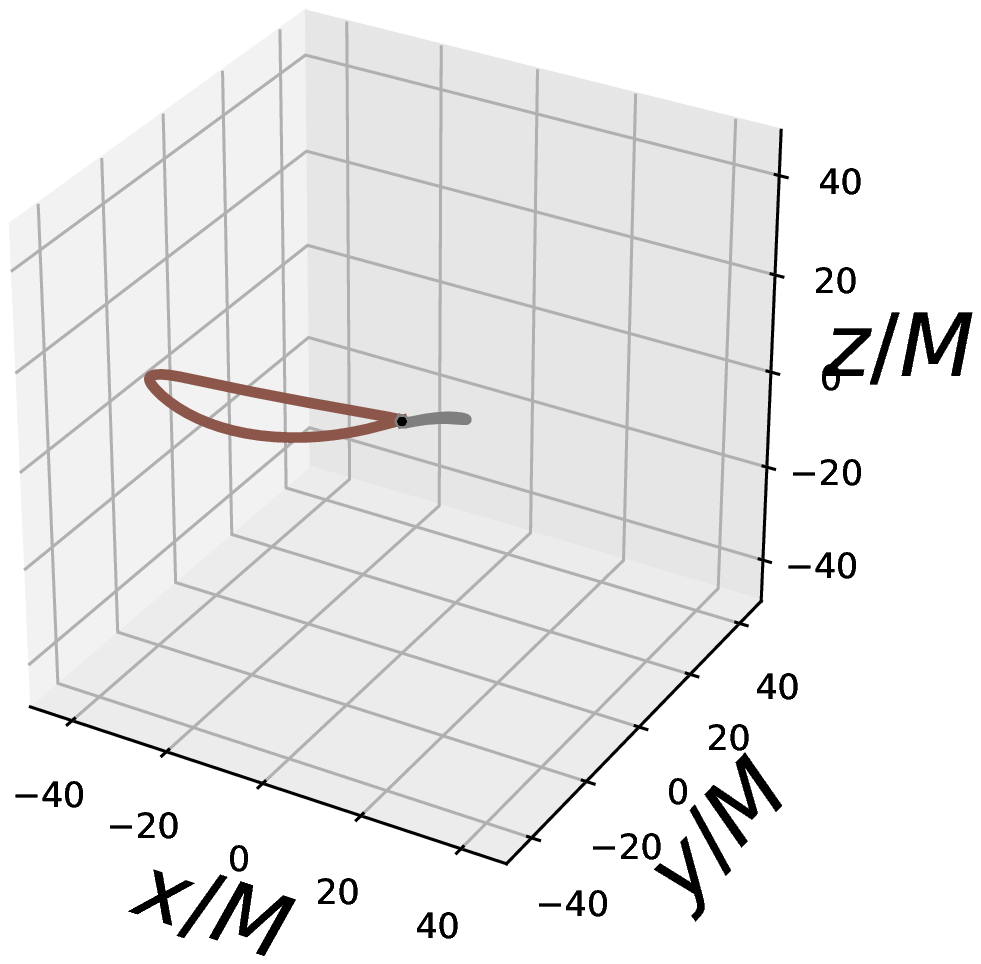}
}\hspace{-1.09cm}\subfloat[\label{fig:swallow}$t\approx5L_{i}$]{\includegraphics[scale=0.5]{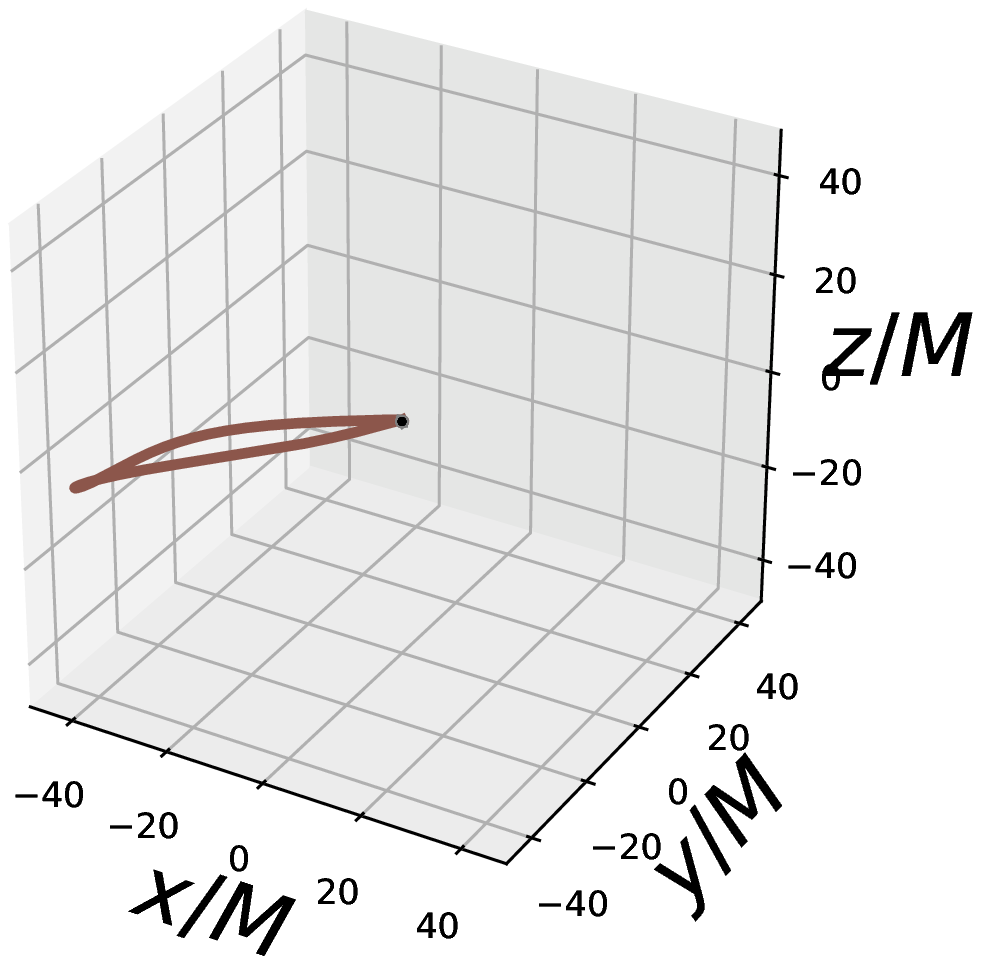}}

\protect\caption{\label{fig:shortening}Snapshots of string configuration of example A at time $t\approx0.9L_{i},2.7L_{i},4L_{i}$
and $5L_{i}$, after the two sub-loops are formed. The brown one has an
angular momentum (almost) in the same direction as the black hole and is rotating
clockwise, whereas the gray one rotates counterclockwise. We can
see that the gray loop gradually turns into a double-line with a decreasing
length, and disappears in the last figure.
On the other hand, for the brown loop, energy extraction
from the black hole spin defeats the energy loss due to horizon friction,
and the length of the loop increases over time.}
\end{figure}

\begin{figure}[!]
\subfloat[$t\approx4.6L_{i}$]{\includegraphics[scale=0.45]{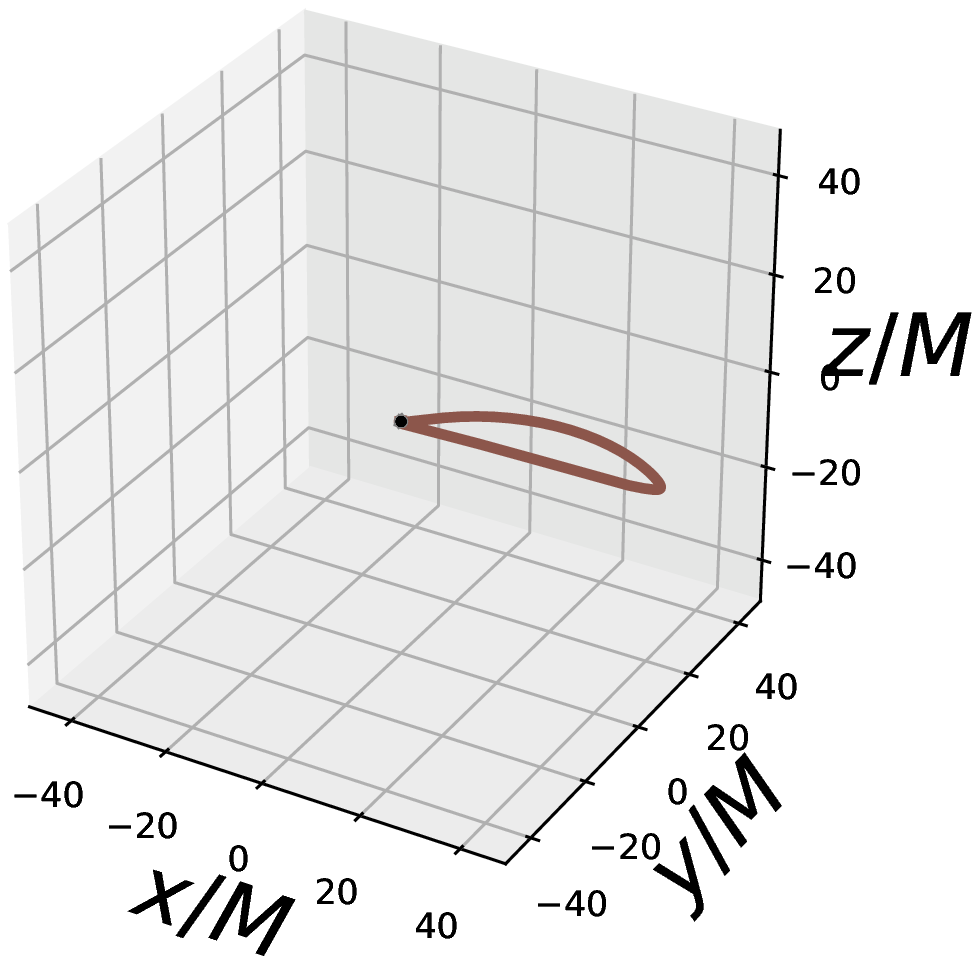}

}\subfloat[$t\approx17.5L_{i}$]{\includegraphics[scale=0.45]{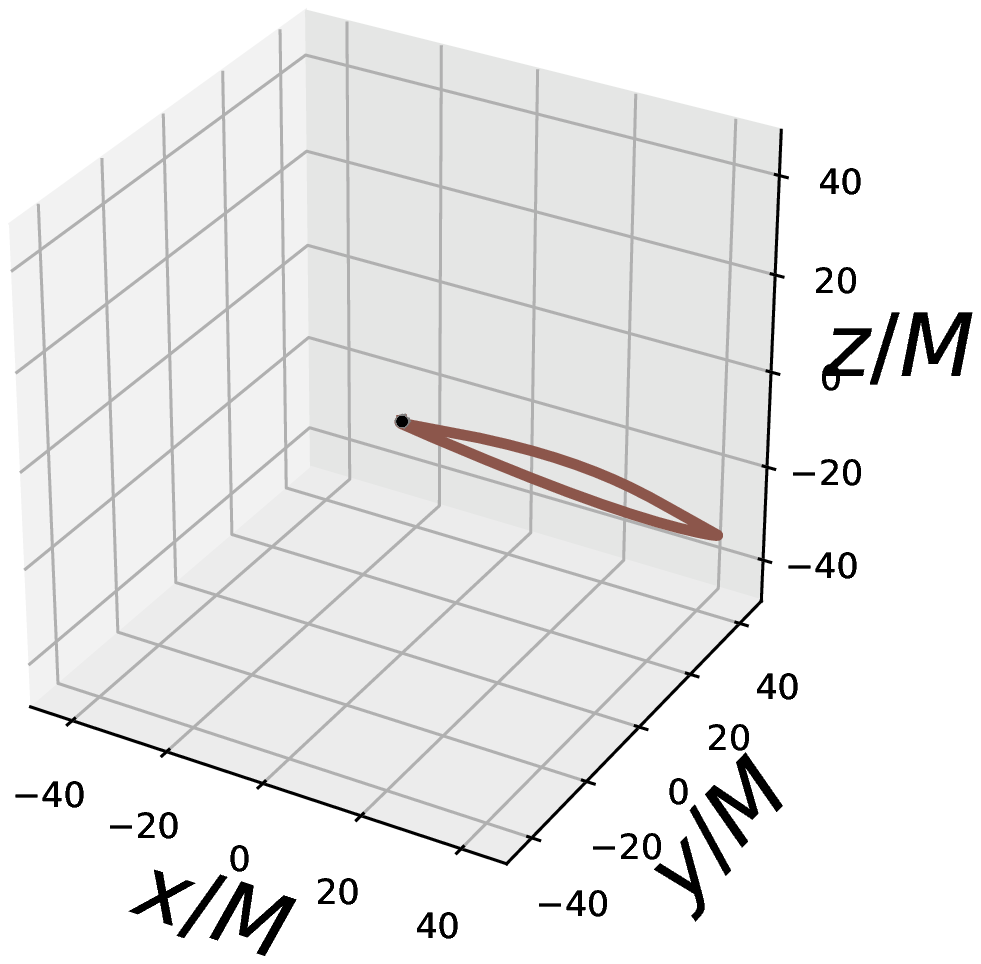}}\subfloat[$t\approx24.8L_{i}$]{\includegraphics[scale=0.45]{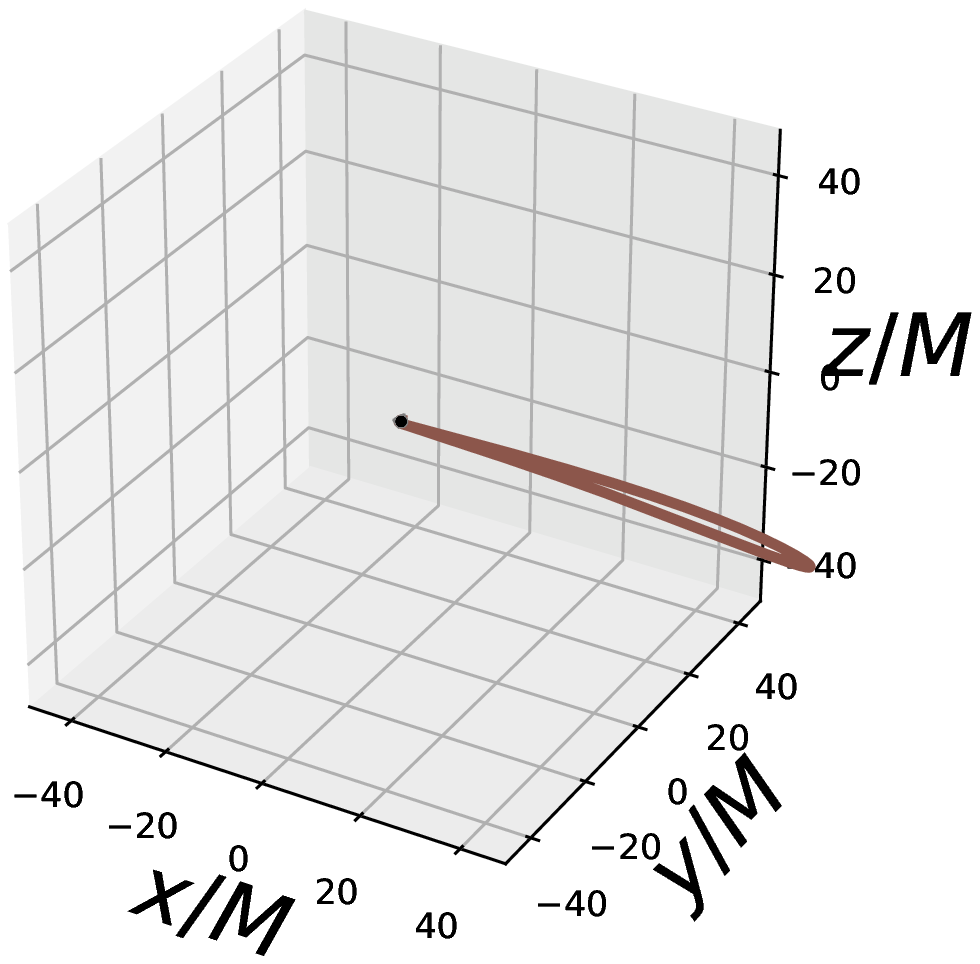}}

\protect\caption{\label{fig:lengthening}Snapshots of string configuration of example A at time
$t\approx4.6L_{i},17.5L_{i}$ and $24.8L_{i}$, after the shortening
loop is completely absorbed. The remaining brown loop rotates (almost) in the
same direction as the black hole. It gradually turns into a double-line
with an ever-increasing length (see fig. \ref{fig:L(t)}).}
\end{figure}

As we can see from fig. \ref{fig:string_before_capture}, in the early stage the string motions are more or less the same in example A and example B (except that the loop is ``tilted'' in example B). However, the difference in black hole spin then leads to different kinds of string evolution. For example A, the sub-loop rotating in the same direction as the black hole spin (clockwise) extracts so much rotational energy that it is able to defy horizon friction, get lengthened, and turn into a growing double-line. The other sub-loop, while also turning into a double-line, would be gradually swallowed. This process is shown with snapshots in figs. \ref{fig:shortening} and \ref{fig:lengthening}.
The simulation begins with a big circle of vanishing velocity, and ends up with an ever-growing, rotating double-line, the tip of which moves at a speed close to the speed of light.

\begin{figure}[!]
\hspace{0.8cm}\subfloat[$t\approx21.9L_{i}$]{\includegraphics[scale=0.45]{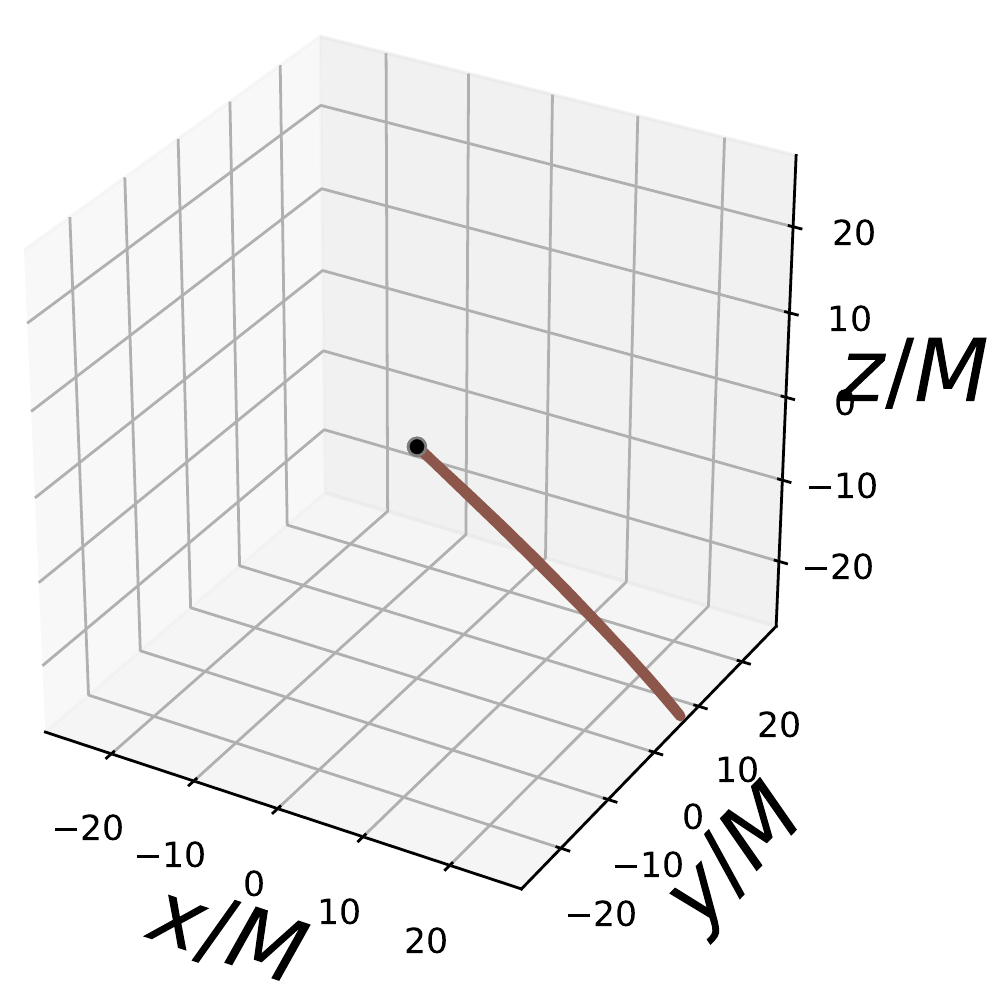}}
\hspace{0.8cm}
\subfloat[$t\approx 54L_{i}$]{\includegraphics[scale=0.45]{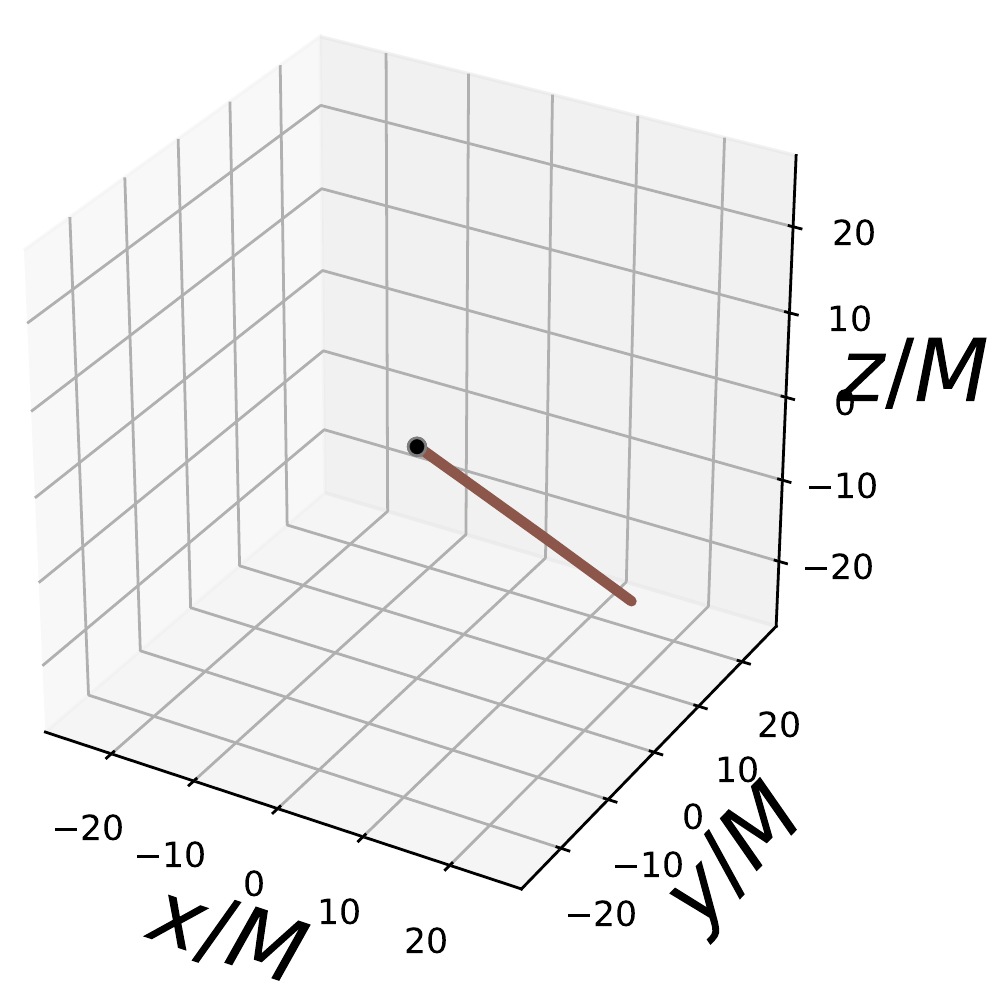}}\hspace{0.8cm}
\subfloat[$t\approx 76.5L_{i}$]{\includegraphics[scale=0.45]{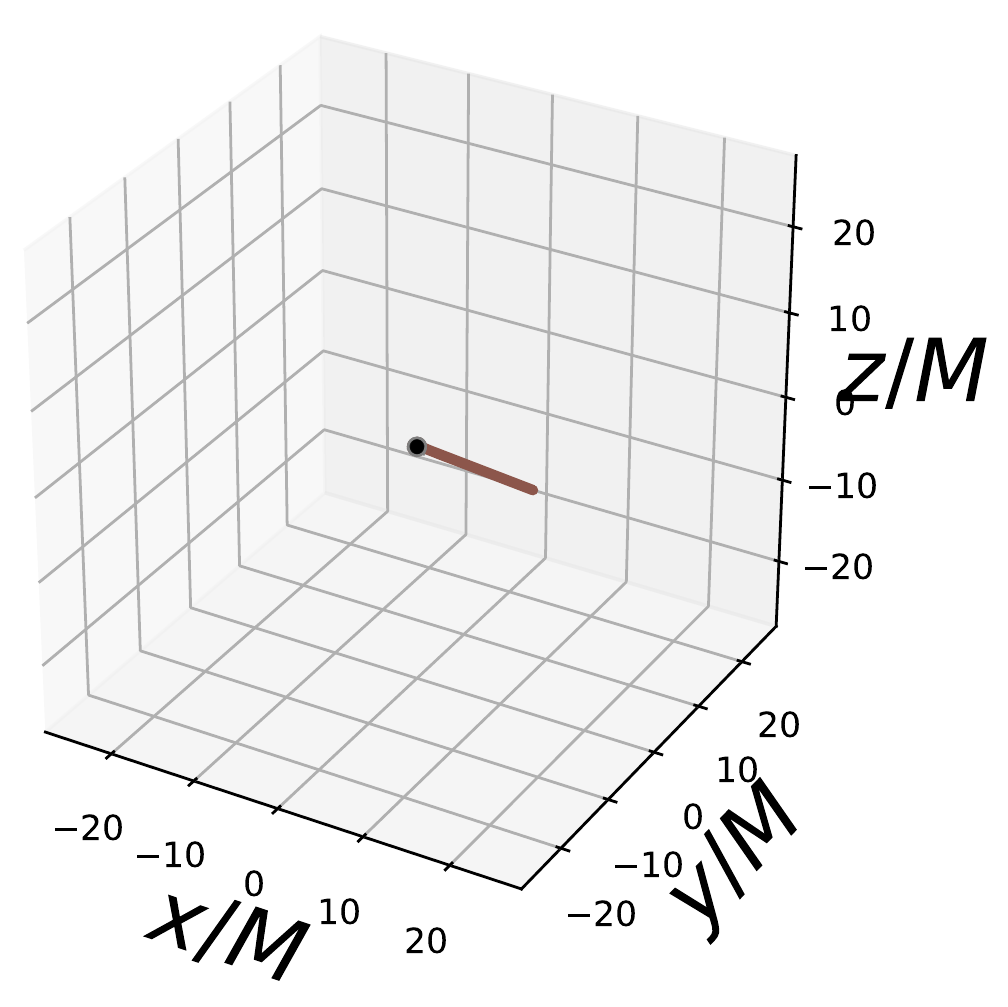}}

\protect\caption{\label{fig:shortening2}Snapshots of string configuration of example B at time
$t\approx21.9L_{i},54L_{i}$ and $76.5L_{i}$, after one sub-loop is completely absorbed. The remaining loop rotates with an angular velocity that has a negative $z$-component. The three snapshots are chosen such that the loop lies almost in the plane of $y=0$. As it turns into a decaying double-line (see also fig. \ref{fig:L(t)2}), the loop is pulled towards the equatorial plane of the black hole.}
\end{figure}

As for example B, due to the relatively small black hole spin, both sub-loops would eventually be absorbed.\footnote{Whether energy extraction can defeat horizon friction depends not only on the black hole spin but also on the string length. A sufficiently large $L_i$ may also lead to string-lengthening even for a small black hole spin. Such an example is not presented here due to the cost of simulating a large string loop.} The one whose angular momentum has a negative $z$-component (the same as the black hole spin) struggles for a longer time before being eaten up as it is able to borrow energy from the black hole. {The evolution of this sub-loop at late times is shown in fig. \ref{fig:shortening2}. It turns into a  shrinking double line whose motion gets closer to the equatorial plane. This outcome agrees with the predictions in ref. \cite{Xing:2020ecz}.}

\subsubsection{Comparing with auxiliary curve}

A goal of the present work is to check whether the auxiliary curve can be used to describe the string motion, especially when the black hole spin is relatively large, and when the string motion is not close to the equatorial plane. The initial configuration of the auxiliary
curve can be obtained from eq. (\ref{eq:auxiliary}), which gives
\begin{equation}
\dot{\boldsymbol{x}}(\sigma,t=0)=\frac{1}{2}\left[-\frac{\text{d}\boldsymbol{\mathsf{a}}(\sigma)}{\text{d}\zeta}+\frac{\text{d}\boldsymbol{\mathsf{a}}(-\sigma)}{\text{d}\zeta}\right],
\end{equation}
\begin{equation}
\boldsymbol{x}^{\prime}(\sigma,t=0)=\frac{1}{2}\left[\frac{\text{d}\boldsymbol{\mathsf{a}}(\sigma)}{\text{d}\zeta}+\frac{\text{d}\boldsymbol{\mathsf{a}}(-\sigma)}{\text{d}\zeta}\right].
\end{equation}
Hence
\begin{equation}
\frac{\text{d}\boldsymbol{\mathsf{a}}(\mp\sigma)}{\text{d}\zeta}=\boldsymbol{x}^{\prime}(\sigma)\pm\dot{\boldsymbol{x}}(\sigma),
\end{equation}
which can be numerically integrated along the arc length $\zeta$
to find $\boldsymbol{\mathsf{a}}(\zeta)$. Then we solve eq. (\ref{eq:eom of auxiliary})
to find the deformation of $\boldsymbol{\mathsf{a}}(\zeta)$.

In order to see how well the auxiliary curve behaves, one can first compare
its length $L_{\mathsf{a}}(t)$ to the length of the simulated string
$L(t)$, where $L(t)$ is obtained according to eq. (\ref{eq:invariant length})
(only string segment away from the horizon should be taken into account).
The comparison tells whether the evolution of the
auxiliary curve correctly accounts for the energy transfer between the
black hole and the string. If the auxiliary curve gives a good approximation,
one should have $L=L_{\mathsf{a}}/2$ for any string configurations
and black hole spins. 

\begin{figure}[!]
\includegraphics[scale=0.35]{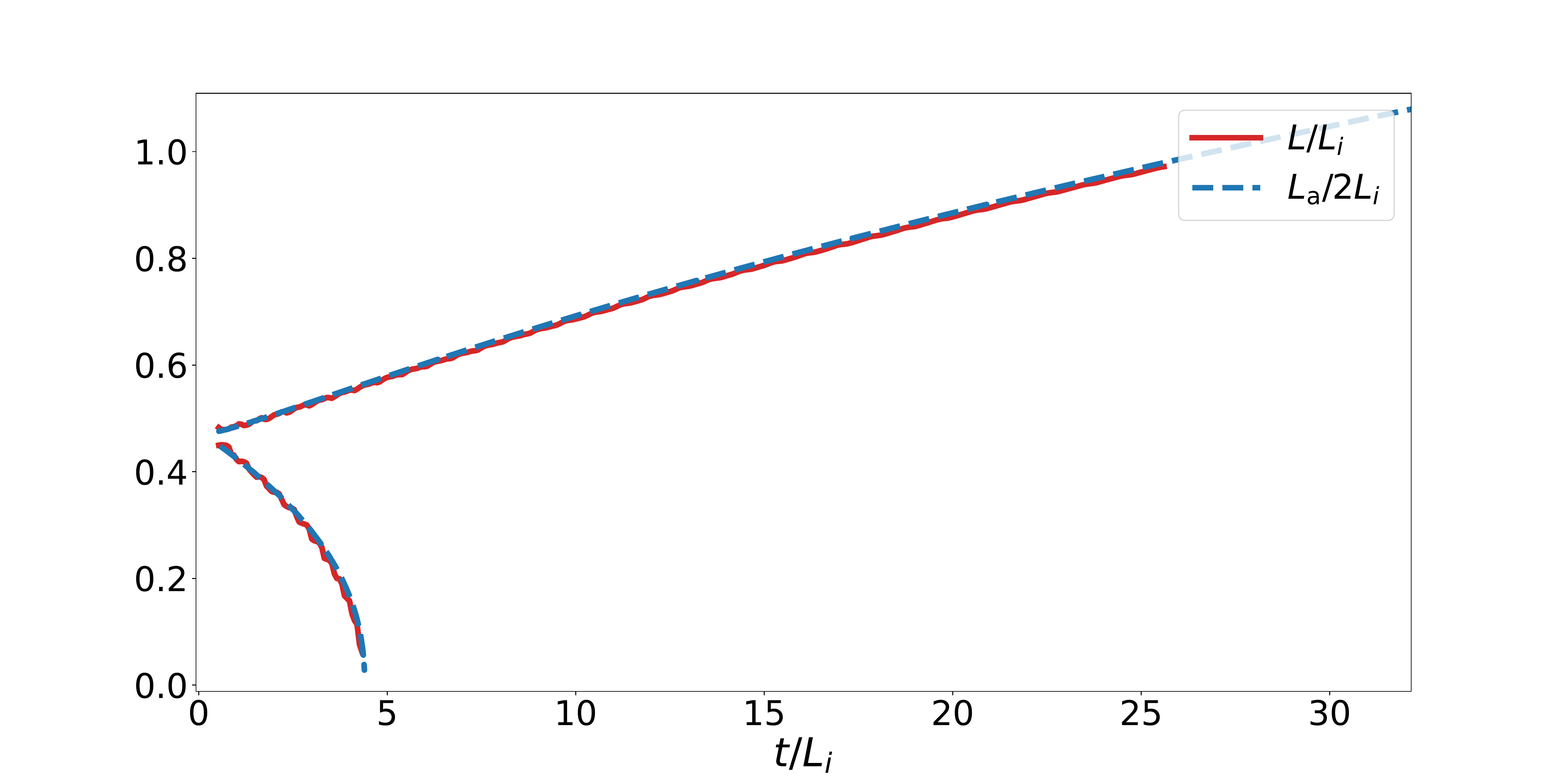}

\caption{\label{fig:L(t)}Example A. String length obtained from the auxiliary curve ($L_{\mathsf{a}}(t)/2$,
dashed blue) and that from actual string simulation ($L(t)$, solid
red). }
\end{figure}

\begin{figure}[!]
\includegraphics[scale=0.35]{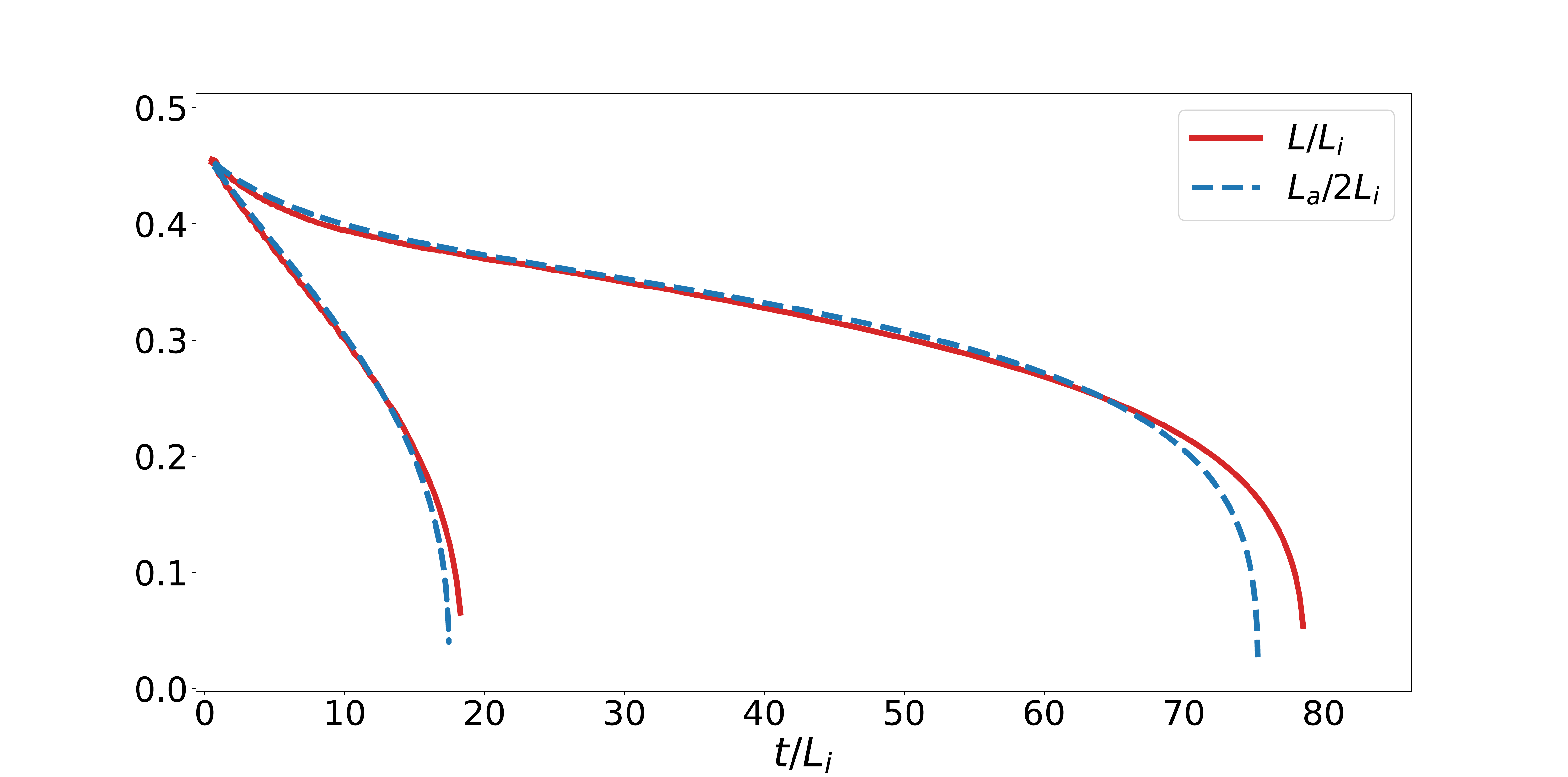}

\caption{\label{fig:L(t)2}Example B. String length obtained from the auxiliary curve ($L_{\mathsf{a}}(t)/2$,
dashed blue) and that from actual string simulation ($L(t)$, solid
red). }
\end{figure}

In our two examples, all sub-loops have string lengths $\sim L_{i}/2$
when they are formed at $t\approx0.5L_{i}$. For example A,  one sub-loop grows afterwards,
while the other decays. From fig. \ref{fig:L(t)}, we can see that
$L(t)$ and $L_{\mathsf{a}}(t)/2$ agree with each other very well. As for example B, both sub-loops are gradually eaten up by the black hole. The comparison of $L(t)$ and $L_{\mathsf{a}}(t)/2$ are shown in fig. \ref{fig:L(t)2}. Except for the discrepancy at late times, the prediction from the auxiliary curve is in general compatible with our experiment.

It should be noted that the good agreement shown in fig. \ref{fig:L(t)} indicates that the horizon friction experienced by the loop in example A is accurately described by the estimated fiction term in eq. (\ref{eq:Q}) or (\ref{eq:eom of auxiliary}); this is understood as a happy coincidence. We found discrepancies for loops not rotating near the black hole's equatorial plane, when the horizon friction may also depend on the black hole spin (see footnote \ref{footnote}), especially if the black hole spin is large. 

Moreover, the consistency in $L$ does not mean the string configurations reconstructed from the auxiliary curve resemble those from simulations. In fact, we found in our examples that deviations in string shape develop with time, which indicates an error in the auxiliary curve equation. We also found that increasing the string length and decreasing the black hole spin alleviate the discrepancy, which is consistent with our discussion in subsec. \ref{predictions}. 

Restricted by the numerical cost of simulating longer strings, we conclude that the auxiliary curve deformation indeed provides a reasonable approximation to the evolution of an actual string. Whether or not it is accurate for sufficiently long strings rotating around a black hole with a large spin remains to be explored. {If the accuracy is shown to be satisfactory for longer strings, then the auxiliary curve approach is much more feasible in studying more complicated string motions.}

\section{Superradiance\label{subsec:Superradiance}}
In the previous section we show two simple examples from string simulations, where the initial string loops are almost circles, but in general a loop can have wiggles composed of a spectrum of modes, and the resulting string motion would be rather complicated. These wiggles could get enhanced by the black hole spin due to the effect of superradiance, which plays a significant role in the string-black hole interaction \cite{Xing:2020ecz}. As a result, self-intersections are expected to be rather common, which can produce a number of free loops if reconnections occur. It is also possible that reconnections rarely take place when two string segments intersect, such as in the model of cosmic super-string. In what follows, we ignore reconnections (as we do in the previous section), and use auxiliary curve deformation to demonstrate superradiance with several examples.

\begin{figure}[!]
\centering
\subfloat[\label{auxiliary_superradiance_0} $t=0$]{\includegraphics[scale=0.45]{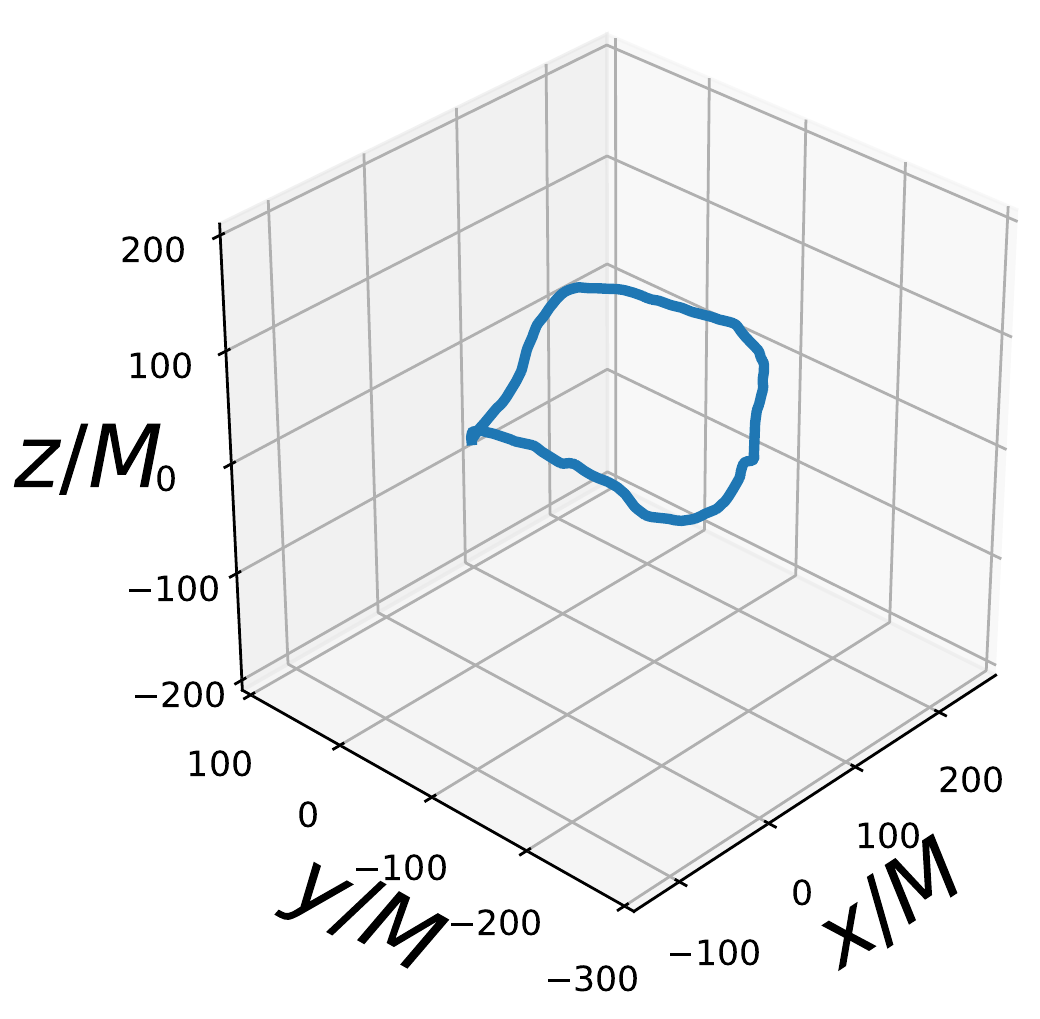}}\hspace{2em}
\subfloat[\label{auxiliary_superradiance_-02} $a=-0.2M$]{\includegraphics[scale=0.45]{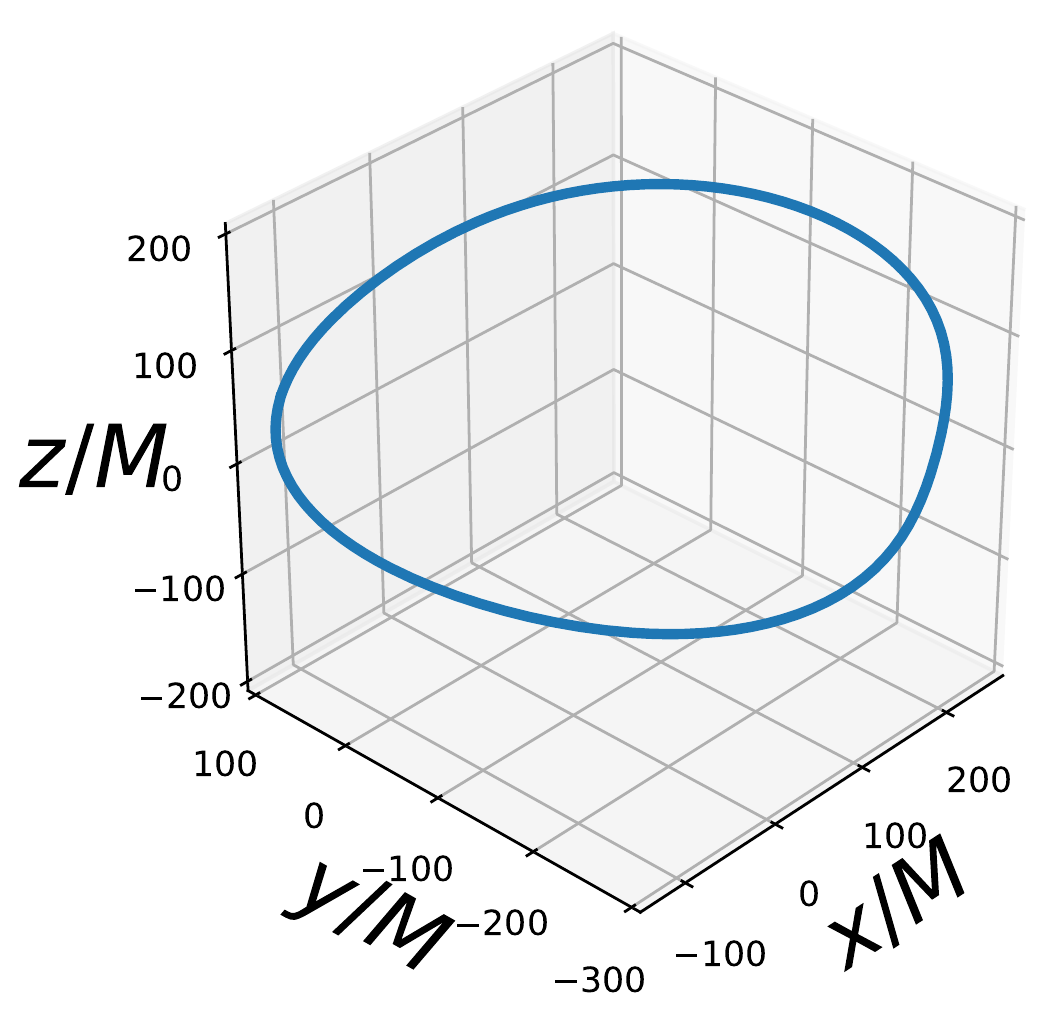}}\hspace{2em}\subfloat[\label{auxiliary_superradiance_-04} $a=-0.4M$]{\includegraphics[scale=0.45]{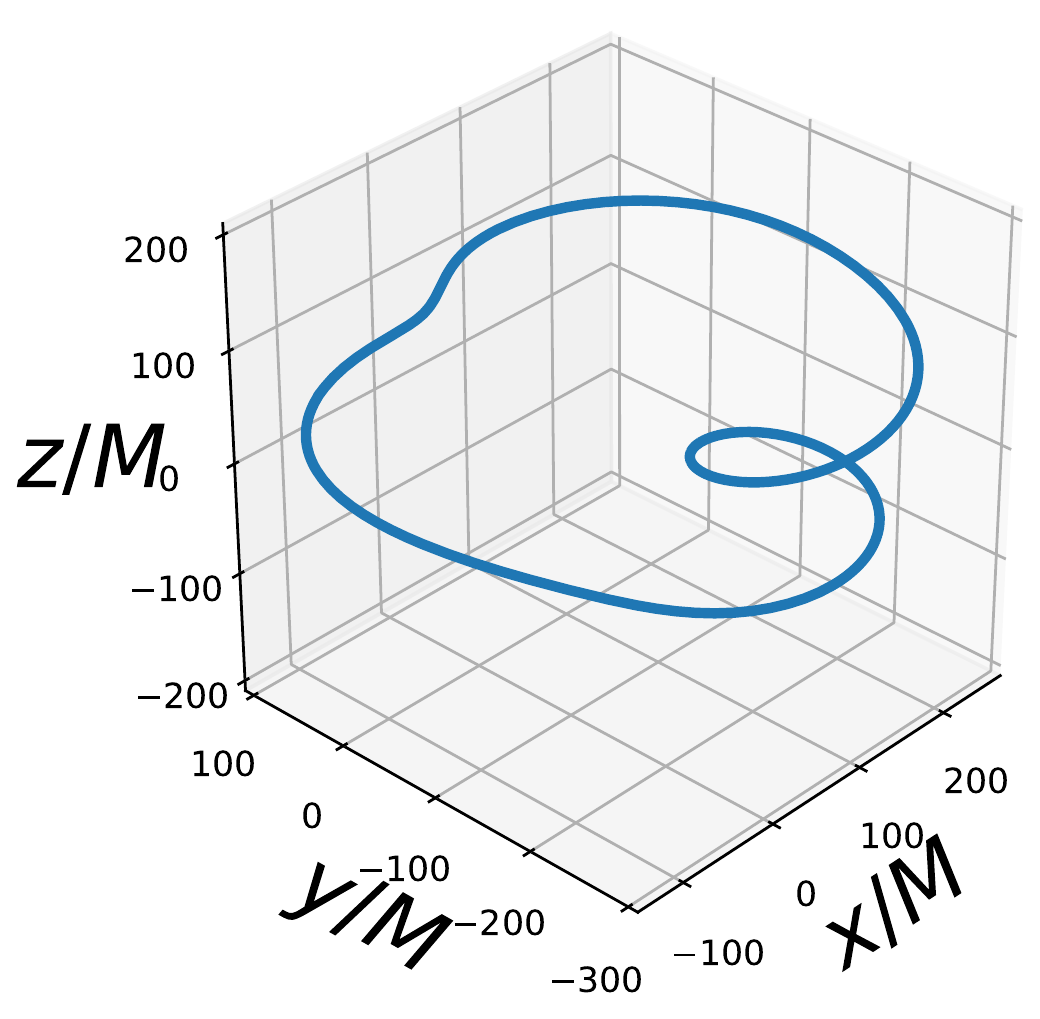}}

\subfloat[\label{auxiliary_superradiance_-06} $a=-0.6M$]{\includegraphics[scale=0.45]{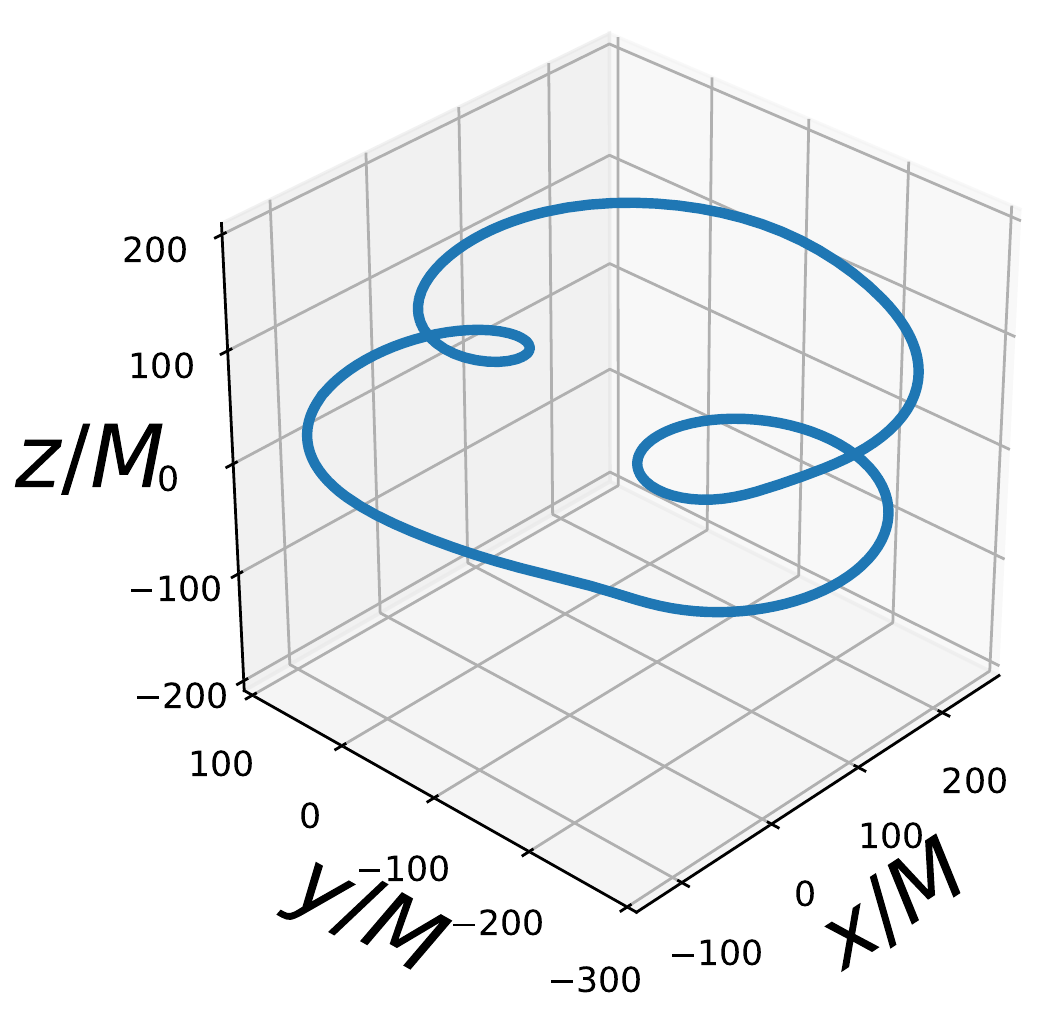}}\hspace{2em}\subfloat[\label{auxiliary_superradiance_-08} $a=-0.8M$]{\includegraphics[scale=0.45]{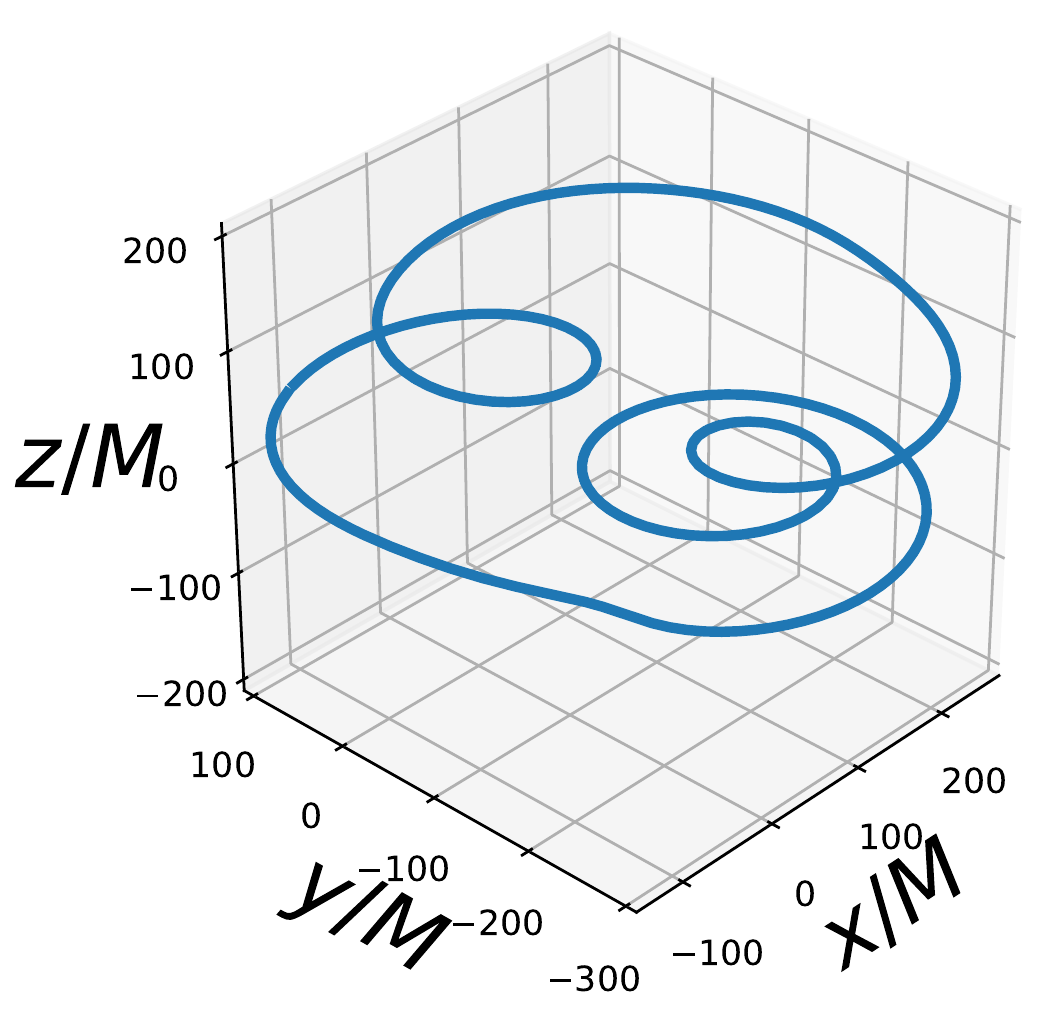}}\hspace{2em}\subfloat[\label{auxiliary_superradiance_08} $a=0.8M$]{\includegraphics[scale=0.45]{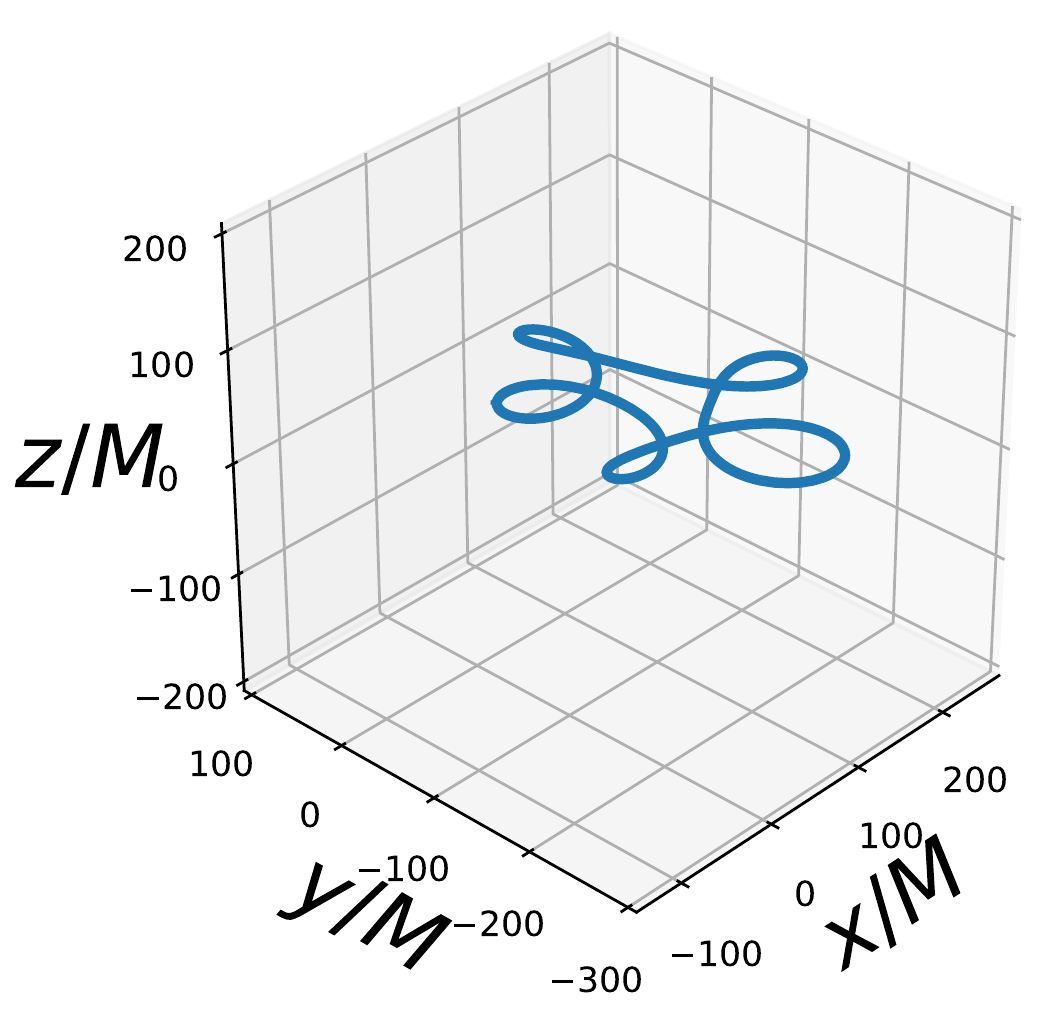}}

\protect\caption{\label{fig:auxiliary_superradiance}(a) A random auxiliary curve with small wiggles. (b-f) Snapshots of the curve after some deformation in the case of different black hole spins characterized by the Kerr parameter $a$.} 
\end{figure}

We first construct a random auxiliary curve $\mathsf{\bold{a}}(\zeta)$, which should satisfy $|\mathsf{\bold{a}}^\prime|=1$. Details of how this can be done numerically are described in ref. \cite{Xing:2020ecz}. This curve is shown in fig. \ref{auxiliary_superradiance_0}. Its length is $L_\mathsf{a}=400M$ and the orientation is counterclockwise from the angle of the figure.

We can then evolve this curve by eq. (\ref{eq:eom of auxiliary}) with different black hole spins. The results are shown in figs. \ref{auxiliary_superradiance_-02}-\ref{auxiliary_superradiance_08}. Each figure shows a snapshot of the curve after some deformation with a specific black hole spin (characterized by the Kerr parameter $a$). Roughly speaking, what the black hole rotation does is filtering out wiggling modes with angular speed larger than that of the black hole \cite{Xing:2020ecz}. In our example, for $a=-0.2M$ (fig. \ref{auxiliary_superradiance_-02}), all wiggles are smoothed out, and the curve simply grows into a circle in the equatorial plane, and the corresponding string turns into an ever-growing double-line rotating clockwise. As $|a|$ increases from $0.4M$ to $0.8M$ (figs. \ref{auxiliary_superradiance_-04}-\ref{auxiliary_superradiance_08}), more and more modes develop and grow. Depending on the black hole spin, a typical auxiliary curve turns into a closed growing coil-like configuration, where the coils are asymptotically parallel to the black hole's equatorial plane. In fig. \ref{auxiliary_superradiance_08} we also show a case with $a=0.8M$. The auxiliary curve first shrinks to a smaller size then re-expands into a growing coil with a helicity opposite to those with $a<0$.

\begin{figure}[!]
\includegraphics[scale=0.35]{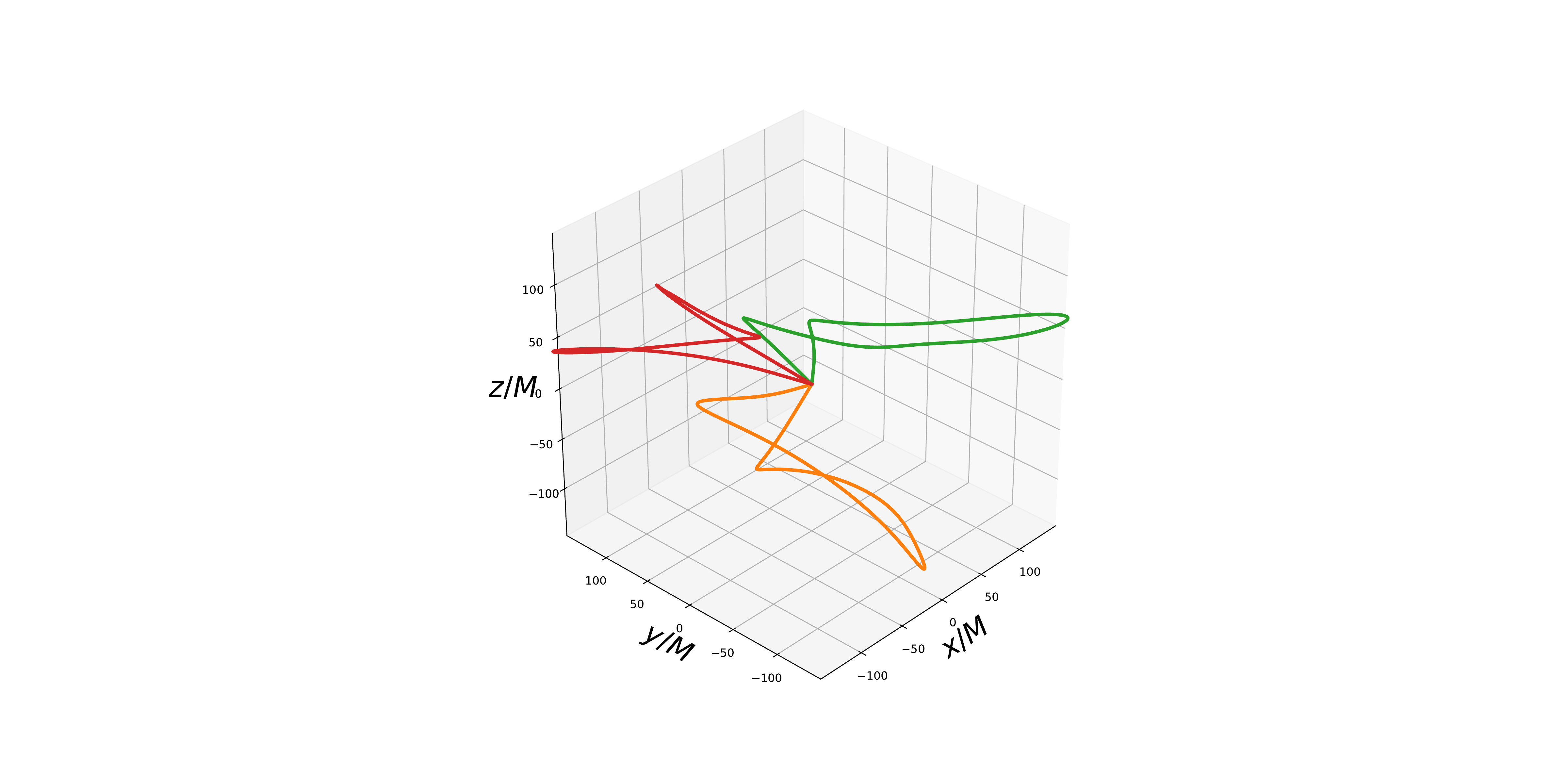}

\caption{\label{fig:superradiance_string}Three snapshots of string loop constructed from the auxiliary curve in fig. \ref{auxiliary_superradiance_-06}.}
\end{figure}

The string configurations corresponding to these coil-like structures are rather interesting. In fig. \ref{fig:superradiance_string} we show several snapshots of the string loop constructed from the auxiliary curve in fig. \ref{auxiliary_superradiance_-06}. We can clearly see three spikes in the loop. The loop would gradually move to the black hole's equatorial plane, with more segments captured by the black hole. The spikes then become rotating double-lines.

Apparently, strings reconstructed from the coils can be rather complicated, where self-intersections certainly occur.  If free loops are generated via string reconnections, we would have a new auxiliary curve that continues deforming, leading to new string motion. Although straightforward in principle, a detailed investigation of how a general auxiliary curve evolves with string reconnections taken into account needs careful numerical work, and is left for future research.

\section{Conclusions and discussion \label{discussion}}
By introducing the auxiliary curve, ref. \cite{Xing:2020ecz} shows that a loop captured by a rotating black hole can experience a complicated motion as it extracts energy from the black hole spin. A number of free loops can be generated as string segments self-intersect and reconnect, and the remaining loop can turn into a number of growing double-lines rotating around the black hole in the equatorial plane, with the tips moving at the speed of light, until the black hole loses its rotational energy.

Using the auxiliary curve deformation to describe the string motion is based on the following assumptions: (1) the string outside the black hole moves as a free string in Minkowski space, and (2) the interaction between the string loop and the black hole is described by the torque in the quasi-stationary string solution with a correction (the ``horizon friction'') due to the slow string motion. These approximations may look reasonable, but they were not justified. 

In this work we have carried out direct numerical simulations to inspect the validity of the auxiliary curve treatment. In our numerical experiments, we observe how a string segment is captured by the black hole, and how the loop turns into decaying or growing double-lines. We found that the energy exchange between the black hole and the string can be described by the auxiliary curve deformation to a reasonable accuracy.

In our simulations, we mainly dealt with strings that are relatively smooth, while a general string is expected to be full of wiggles, and the effect of superradiance should be present. In order to fully understand the consequence of superradiance and string reconnections, as well as what kinds of astrophysical/cosmological phenomena they could produce, further numerical work is needed. However, {it would be extremely computationally expensive to do this with direct simulations; solving for the evolution of the auxiliary curve is more feasible. Our results give confidence that the latter approach can be reliable.}

\section*{Acknowledgments}
HD was supported by the U.S. Department of Energy, Office of High Energy Physics, under Award No. de-sc0019470 at Arizona State University, and the NSF NANOGrav Physics Frontier Center No. 2020265. YL was supported by the Simons Investigator Grant 827103. AV was supported by the National Science Foundation under grant No. 2110466. Part of this research was undertaken at Flatiron Institute, enabled by HD's visit and YL's part-time appointment. 
The Flatiron Institute is a division of the Simons Foundation, supported through the generosity of Marilyn and Jim Simons.

\subsection*{Appendix A: deformation of smooth auxiliary curves \label{subsec:String-evolution-from}}

In this appendix we consider three limits of the auxiliary curve equation (eq. (\ref{eq:eom of auxiliary})).
For simplicity, suppose the curve $\boldsymbol{\mathsf{a}}(\zeta)$
is sufficiently ``smooth'' such that superradiance does not occur.

\begin{figure}[!]
\centering
\subfloat[\label{fig:limit1} $a=0$]{\includegraphics[scale=0.26]{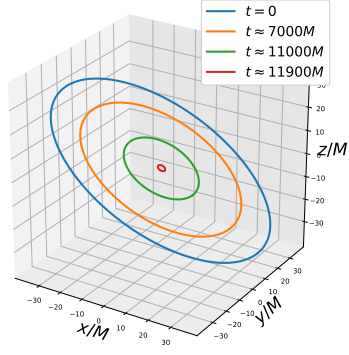}}\subfloat[\label{fig:limit2} $a=-0.8M$]{\includegraphics[scale=0.26]{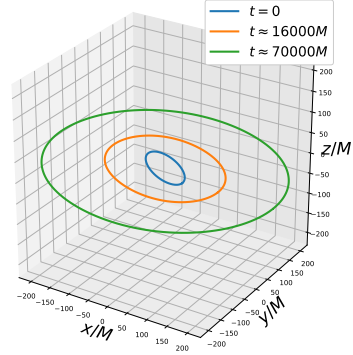}}

\subfloat[\label{fig:limit3a} $a=0.4M$]{\includegraphics[scale=0.26]{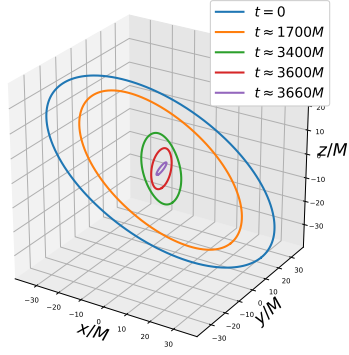}}\subfloat[\label{fig:limit3b} $a=0.8M$]{\includegraphics[scale=0.26]{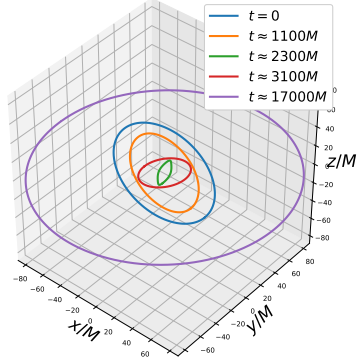}}

\caption{\label{fig:auxiliary limit}Evolution of a smooth auxiliary curve in several typical scenarios. Four sub-figures have the same initial curve (blue): an ellipse with semi-major axis $50M$ and semi-minor
axis $40M$, tilted by $\pi/6$ with respect to the $x\text{-}y$
plane, orientation being counterclockwise. (a) $\boldsymbol{J}=0$,
hence Schwarzschild black hole. The curve gradually shrinks to a small
circle, then a dot in the end. (b) $\boldsymbol{J}=-0.8M^2\hat{\boldsymbol{z}}$,
hence $\boldsymbol{\mathsf{v}}$ points away from the center. The
curve expands into an ever-growing circle perpendicular to the $z$-axis. (c) $\boldsymbol{J}=0.4M^2\hat{\boldsymbol{z}}$, hence $\boldsymbol{\mathsf{v}}$
points inwards. The curve is tilted towards
the $z$-axis, and shrinks into a dot in the meantime. (d) $\boldsymbol{J}=0.8M^2\hat{\boldsymbol{z}}$,
hence $\boldsymbol{\mathsf{v}}$ points inwards. The curve gets tilted, shrinks,
but then its orientation is flipped (red and purple) and it re-expands
into an ever-growing circle perpendicular to the $z$-axis.}

\end{figure}

(1) Firstly, for a non-rotating black hole, we have
$\boldsymbol{\mathsf{v}}\propto\boldsymbol{\mathsf{a}}^{\prime\prime}$.
This is known in geometry as the equation of a ``curve-shortening
flow'' (see, e.g., refs. \cite{gage1984curve, gage1986shrinking,grayson1989shortening}). Asymptotically, any shape of a closed curve
would turn into a shrinking planar circle, and disappear in a dot
after a finite time. An example in this scenario is shown in fig. \ref{fig:limit1}. By eq. (\ref{eq:auxiliary}), a circular auxiliary curve of length
$L_{\mathsf{a}}=2L$ corresponds to a string loop that extends radially
from the black hole to radius $L/\pi$ and then traces the same radial
line back to the black hole. This double-line rotates around
the black hole with angular velocity $\pi/L$, and its tip
moves at the speed of light. Physically, this tells us that, for
a non-rotating black hole, any captured string loop would gradually
turn into a double-line, which would eventually
be eaten up by the black hole. It can easily be shown
that a double-line with length $L$ decays as
\begin{equation}
\dot{L}=-8\pi^{2}\left(\frac{M}{L}\right)^{2}.
\end{equation}

(2) For the second limit, we consider a rotating black hole with angular
momentum $J=|\boldsymbol{J}|\sim M^{2}$ ($M^2$ is the largest angular
momentum a black hole can have). For a simple auxiliary curve, $|\boldsymbol{\mathsf{a}}^{\prime\prime}|\sim1/L$
(which is the curvature of the curve) and $|\boldsymbol{J}\times\boldsymbol{\mathsf{a}}^{\prime}|\sim M^{2}$.
Since $L\gg M$, this means the first term in eq. (\ref{eq:eom of auxiliary})
can be neglected. In this case, if $\boldsymbol{\mathsf{v}}\propto\boldsymbol{J}\times\boldsymbol{\mathsf{a}}^{\prime}$
points outwards, i.e., the curve orientation is opposite to the black
hole rotation, the auxiliary curve expands in the direction perpendicular
to the black hole spin, and becomes increasingly circular (fig. \ref{fig:limit2}). Correspondingly,
the double-line rotates around the black hole, approaching the equatorial plane of the black hole, and the string length increases with
time. It can be shown that a double-line with length $L$ grows as
\begin{equation}
\dot{L}=2\pi\frac{J}{ML}.\label{eq:auxiliary curve lengthening}
\end{equation}
From the above two equations, we can see that whether there is energy
and angular momentum exchanges between the black hole and the captured
string loop depends on the relation between the black hole spin and
the string length. A double-line with length $L$ larger (smaller)
than $4\pi M^{3}/J$ would grow (decay).

(3) As for the third limit, let us consider a case similar to that
in the second limit, except that $\boldsymbol{\mathsf{v}}$
now points inwards. In this case, if the initial curve is not exactly
in the equatorial plane of the black hole, it gets even more tilted towards
the $z$-axis as the curve deforms and shrinks. If the friction term in eq. (\ref{eq:eom of auxiliary}) dominates afterwards, the
auxiliary curve simply shrinks to a dot as in the first limit (fig. \ref{fig:limit3a}). It
is also possible that the curve is tilted such that its orientation
gets flipped. If the string length is sufficiently large, the curve
could start to expand as in the second limit (fig. \ref{fig:limit3a}).

In fig. \ref{fig:auxiliary limit} we show the evolution of auxiliary
curves in these different scenarios. All four sub-figures have the same
initial curve: an ellipse with semi-major axis $50M$ and semi-minor
axis $40M$, tilted by $\pi/6$ with respect to the $x\text{-}y$
plane, its orientation being counterclockwise. The black
hole's angular momentum is parallel to the $z$-axis,
and is given by $\boldsymbol{J}=aM\hat{\boldsymbol{z}}$. We can see how different
black hole spins lead to different outcomes.

\subsection*{Appendix B: initial conditions of simulations \label{subsec:BF}}
In this appendix we show the details of setting up the string loop's initial conditions in our simulations. We would like to transform known string solutions $\boldsymbol{X}(\sigma, t)$ in Minkowski spacetime into $\boldsymbol{x}(\sigma)$ and $\dot{\boldsymbol{x}}(\sigma)$ in Kerr spacetime. 

For notational simplicity, we adopt a general stationary metric (instead of the specific Kerr metric):
\begin{equation}
\text{d}s^{2}=\left(g_{00}(\boldsymbol{x})-f(\boldsymbol{x})\right)\text{d}t^{2}+g(\boldsymbol{x})\text{d}\boldsymbol{x}^{2}+f(\boldsymbol{x})\left(\text{d}t+A_{i}(\boldsymbol{x})\text{d}x^{i}\right)^{2}. \label{eq:appendix metric}
\end{equation}
Comparing with the quasi-isotropic metric (\ref{eq:iso}) gives
\begin{align}
g_{00} & = -\left(1 - \frac{2Mr}{\rho^2} \right),\\
f & = \left(\frac{2Mr}{\rho^2}\right)^2\left(1+\frac{2Mr}{\rho^2}\right)^{-1}, \\
g & = \left(\frac{\rho}{\bar{r}}\right)^2,\\
A_1 & = \left(1+\frac{\rho^2}{2Mr}\right)\frac{ay}{\bar{r}^2} ,\\ 
A_2 & = -\left(1+\frac{\rho^2}{2Mr}\right)\frac{ax}{\bar{r}^2} ,\\ 
A_3 & = 0.
\end{align}

Suppose the initial conditions of the string loop in simulations are given by
\begin{align}
\boldsymbol{x} & =\boldsymbol{X},\label{eq:from X to x2}\\
\dot{\boldsymbol{x}} & = F\left(\dot{\boldsymbol{X}}+B\boldsymbol{X}^{\prime}\right),\label{eq:from Xdot to xdot2}
\end{align}
where $B$ and $F$ are functions of $\boldsymbol{X},\dot{\boldsymbol{X}}$
and $\boldsymbol{X}^{\prime}$. The task is to find expressions for $B$ and $F$ such that the string obeys the following two conditions (eqs. (\ref{eq:gauge}) and (\ref{eq:def of F}))
\begin{equation}
\dot{x}\cdot x^{\prime}=0,
\end{equation}
\begin{equation}
\dot{x}^{2}+F^{2}x^{\prime2}=0.
\end{equation}
Using the metric (\ref{eq:appendix metric}) as well as the fact that the Minkowski solution $\boldsymbol{X}$ satisfies $\dot{\boldsymbol{X}}\cdot\boldsymbol{X}^{\prime} =0$ and $\dot{\boldsymbol{X}}^{2}+\boldsymbol{X}^{\prime2} =1$, these two conditions can be spelled out as equations for $B$ and $F$:
\begin{equation}
\left[g\boldsymbol{X}^{\prime 2} + f\left(\boldsymbol{A}\cdot \boldsymbol{X}^{\prime}\right)^2\right]B+f\left(\boldsymbol{A}\cdot \boldsymbol{X}^{\prime}\right)\left(F^{-1} + \boldsymbol{A}\cdot \dot{\boldsymbol{X}}\right)=0,\label{eq:BF1}
\end{equation}
\begin{equation}
\left\{f\left[\left(\boldsymbol{A}\cdot \dot{\boldsymbol{X}}\right)^2 +\left(\boldsymbol{A}\cdot \boldsymbol{X}^{\prime}\right)^2+\left(\boldsymbol{A}\cdot \boldsymbol{X}^{\prime}\right)\left(\boldsymbol{A}\cdot \dot{\boldsymbol{X}}\right)B\right]+g\right\}F^2+f\left[2\left(\boldsymbol{A}\cdot \dot{\boldsymbol{X}}\right)+\left(\boldsymbol{A}\cdot \boldsymbol{X}^{\prime}\right)B\right]F + g_{00}=0.
\end{equation}
From the first equation we can write $B$ as a function of $F$. Inserting it into the second equation gives a quadratic equation for $F$:
\begin{equation}
\left[\frac{fg\left(\boldsymbol{A}\cdot\dot{\boldsymbol{X}}\right)^{2}\boldsymbol{X}^{\prime2}}{g\boldsymbol{X}^{\prime2}+f\left(\boldsymbol{A}\cdot\boldsymbol{X}^{\prime}\right)^{2}}+f\left(\boldsymbol{A}\cdot\boldsymbol{X}^{\prime}\right)^{2}+g\right]F^{2}+\frac{2fg\left(\boldsymbol{A}\cdot\dot{\boldsymbol{X}}\right)\boldsymbol{X}^{\prime2}}{g\boldsymbol{X}^{\prime2}+f\left(\boldsymbol{A}\cdot\boldsymbol{X}^{\prime}\right)^{2}}F+\left[g_{00}-\frac{f^{2}\left(\boldsymbol{A}\cdot\boldsymbol{X}^{\prime}\right)^{2}}{g\boldsymbol{X}^{\prime2}+f\left(\boldsymbol{A}\cdot\boldsymbol{X}^{\prime}\right)^{2}}\right]=0.
\end{equation}
It can easily be checked that, if this equation has real solutions, the two roots are of different signs. By the definition of $F$ (eq. (\ref{eq:def of F})), $F\geq 0$. So we should take the positive root, and $F(\sigma)$ along the string is thus determined. Then we use eq. (\ref{eq:BF1}) to obtain $B(\sigma)$. 

The existence of solutions for $B$ and $F$ is not guaranteed if a string segment is too close to the black hole at the initial time, because the quadratic equation for $F$ may not have real solutions. Empirically, in our numerical experiments, we found that solutions exist if the string point closest to the black hole is farther than $\mathcal{O}(1\text{-}10)$ times the horizon.

\bibliography{StringBHbib}

\begin{thebibliography}{27}
\expandafter\ifx\csname natexlab\endcsname\relax\def\natexlab#1{#1}\fi
\expandafter\ifx\csname bibnamefont\endcsname\relax
  \def\bibnamefont#1{#1}\fi
\expandafter\ifx\csname bibfnamefont\endcsname\relax
  \def\bibfnamefont#1{#1}\fi
\expandafter\ifx\csname citenamefont\endcsname\relax
  \def\citenamefont#1{#1}\fi
\expandafter\ifx\csname url\endcsname\relax
  \def\url#1{\texttt{#1}}\fi
\expandafter\ifx\csname urlprefix\endcsname\relax\def\urlprefix{URL }\fi
\providecommand{\bibinfo}[2]{#2}
\providecommand{\eprint}[2][]{\url{#2}}

\bibitem[{\citenamefont{Kibble}(1976)}]{Kibble:1976sj}
\bibinfo{author}{\bibfnamefont{T.~W.~B.} \bibnamefont{Kibble}},
  \bibinfo{journal}{J. Phys. A} \textbf{\bibinfo{volume}{9}},
  \bibinfo{pages}{1387} (\bibinfo{year}{1976}).

\bibitem[{\citenamefont{Vilenkin and Shellard}(2000)}]{Vilenkin:2000jqa}
\bibinfo{author}{\bibfnamefont{A.}~\bibnamefont{Vilenkin}} \bibnamefont{and}
  \bibinfo{author}{\bibfnamefont{E.~P.~S.} \bibnamefont{Shellard}},
  \emph{\bibinfo{title}{{Cosmic Strings and Other Topological Defects}}}
  (\bibinfo{publisher}{Cambridge University Press}, \bibinfo{year}{2000}), ISBN
  \bibinfo{isbn}{978-0-521-65476-0}.

\bibitem[{\citenamefont{Damour and Vilenkin}(2000)}]{Damour:2000wa}
\bibinfo{author}{\bibfnamefont{T.}~\bibnamefont{Damour}} \bibnamefont{and}
  \bibinfo{author}{\bibfnamefont{A.}~\bibnamefont{Vilenkin}},
  \bibinfo{journal}{Phys. Rev. Lett.} \textbf{\bibinfo{volume}{85}},
  \bibinfo{pages}{3761} (\bibinfo{year}{2000}), \eprint{gr-qc/0004075}.

\bibitem[{\citenamefont{Damour and Vilenkin}(2005)}]{Damour:2004kw}
\bibinfo{author}{\bibfnamefont{T.}~\bibnamefont{Damour}} \bibnamefont{and}
  \bibinfo{author}{\bibfnamefont{A.}~\bibnamefont{Vilenkin}},
  \bibinfo{journal}{Phys. Rev. D} \textbf{\bibinfo{volume}{71}},
  \bibinfo{pages}{063510} (\bibinfo{year}{2005}), \eprint{hep-th/0410222}.

\bibitem[{\citenamefont{Vilenkin}(1981)}]{Vilenkin:1981bx}
\bibinfo{author}{\bibfnamefont{A.}~\bibnamefont{Vilenkin}},
  \bibinfo{journal}{Phys. Lett. B} \textbf{\bibinfo{volume}{107}},
  \bibinfo{pages}{47} (\bibinfo{year}{1981}).

\bibitem[{\citenamefont{Buchmuller et~al.}(2020)\citenamefont{Buchmuller,
  Domcke, and Schmitz}}]{Buchmuller:2020lbh}
\bibinfo{author}{\bibfnamefont{W.}~\bibnamefont{Buchmuller}},
  \bibinfo{author}{\bibfnamefont{V.}~\bibnamefont{Domcke}}, \bibnamefont{and}
  \bibinfo{author}{\bibfnamefont{K.}~\bibnamefont{Schmitz}},
  \bibinfo{journal}{Phys. Lett. B} \textbf{\bibinfo{volume}{811}},
  \bibinfo{pages}{135914} (\bibinfo{year}{2020}), \eprint{2009.10649}.

\bibitem[{\citenamefont{Ellis and Lewicki}(2021)}]{Ellis:2020ena}
\bibinfo{author}{\bibfnamefont{J.}~\bibnamefont{Ellis}} \bibnamefont{and}
  \bibinfo{author}{\bibfnamefont{M.}~\bibnamefont{Lewicki}},
  \bibinfo{journal}{Phys. Rev. Lett.} \textbf{\bibinfo{volume}{126}},
  \bibinfo{pages}{041304} (\bibinfo{year}{2021}), \eprint{2009.06555}.

\bibitem[{\citenamefont{Blanco-Pillado
  et~al.}(2021)\citenamefont{Blanco-Pillado, Olum, and
  Wachter}}]{Blanco-Pillado:2021ygr}
\bibinfo{author}{\bibfnamefont{J.~J.} \bibnamefont{Blanco-Pillado}},
  \bibinfo{author}{\bibfnamefont{K.~D.} \bibnamefont{Olum}}, \bibnamefont{and}
  \bibinfo{author}{\bibfnamefont{J.~M.} \bibnamefont{Wachter}},
  \bibinfo{journal}{Phys. Rev. D} \textbf{\bibinfo{volume}{103}},
  \bibinfo{pages}{103512} (\bibinfo{year}{2021}), \eprint{2102.08194}.

\bibitem[{\citenamefont{Hindmarsh and Kume}(2022)}]{Hindmarsh:2022awe}
\bibinfo{author}{\bibfnamefont{M.}~\bibnamefont{Hindmarsh}} \bibnamefont{and}
  \bibinfo{author}{\bibfnamefont{J.}~\bibnamefont{Kume}}
  (\bibinfo{year}{2022}), \eprint{2210.06178}.

\bibitem[{\citenamefont{Lonsdale and Moss}(1988)}]{Lonsdale:1988xd}
\bibinfo{author}{\bibfnamefont{S.}~\bibnamefont{Lonsdale}} \bibnamefont{and}
  \bibinfo{author}{\bibfnamefont{I.}~\bibnamefont{Moss}},
  \bibinfo{journal}{Nucl. Phys. B} \textbf{\bibinfo{volume}{298}},
  \bibinfo{pages}{693} (\bibinfo{year}{1988}).

\bibitem[{\citenamefont{De~Villiers and Frolov}(1998)}]{DeVilliers:1997nk}
\bibinfo{author}{\bibfnamefont{J.-P.} \bibnamefont{De~Villiers}}
  \bibnamefont{and} \bibinfo{author}{\bibfnamefont{V.~P.}
  \bibnamefont{Frolov}}, \bibinfo{journal}{Int. J. Mod. Phys. D}
  \textbf{\bibinfo{volume}{7}}, \bibinfo{pages}{957} (\bibinfo{year}{1998}),
  \eprint{gr-qc/9711045}.

\bibitem[{\citenamefont{De~Villiers and Frolov}(1999)}]{DeVilliers:1998nm}
\bibinfo{author}{\bibfnamefont{J.-P.} \bibnamefont{De~Villiers}}
  \bibnamefont{and} \bibinfo{author}{\bibfnamefont{V.~P.}
  \bibnamefont{Frolov}}, \bibinfo{journal}{Class. Quant. Grav.}
  \textbf{\bibinfo{volume}{16}}, \bibinfo{pages}{2403} (\bibinfo{year}{1999}),
  \eprint{gr-qc/9812016}.

\bibitem[{\citenamefont{Frolov and Larsen}(1999)}]{Frolov:1999pj}
\bibinfo{author}{\bibfnamefont{A.~V.} \bibnamefont{Frolov}} \bibnamefont{and}
  \bibinfo{author}{\bibfnamefont{A.~L.} \bibnamefont{Larsen}},
  \bibinfo{journal}{Class. Quant. Grav.} \textbf{\bibinfo{volume}{16}},
  \bibinfo{pages}{3717} (\bibinfo{year}{1999}), \eprint{gr-qc/9908039}.

\bibitem[{\citenamefont{Snajdr and Frolov}(2003)}]{Snajdr:2002aa}
\bibinfo{author}{\bibfnamefont{M.}~\bibnamefont{Snajdr}} \bibnamefont{and}
  \bibinfo{author}{\bibfnamefont{V.~P.} \bibnamefont{Frolov}},
  \bibinfo{journal}{Class. Quant. Grav.} \textbf{\bibinfo{volume}{20}},
  \bibinfo{pages}{1303} (\bibinfo{year}{2003}), \eprint{gr-qc/0211018}.

\bibitem[{\citenamefont{Dubath et~al.}(2007)\citenamefont{Dubath,
  Sakellariadou, and Viallet}}]{Dubath:2006vs}
\bibinfo{author}{\bibfnamefont{F.}~\bibnamefont{Dubath}},
  \bibinfo{author}{\bibfnamefont{M.}~\bibnamefont{Sakellariadou}},
  \bibnamefont{and} \bibinfo{author}{\bibfnamefont{C.~M.}
  \bibnamefont{Viallet}}, \bibinfo{journal}{Int. J. Mod. Phys. D}
  \textbf{\bibinfo{volume}{16}}, \bibinfo{pages}{1311} (\bibinfo{year}{2007}),
  \eprint{gr-qc/0609089}.

\bibitem[{\citenamefont{Frolov et~al.}(1989)\citenamefont{Frolov, Skarzhinsky,
  Zelnikov, and Heinrich}}]{Frolov:1988zn}
\bibinfo{author}{\bibfnamefont{V.~P.} \bibnamefont{Frolov}},
  \bibinfo{author}{\bibfnamefont{V.}~\bibnamefont{Skarzhinsky}},
  \bibinfo{author}{\bibfnamefont{A.}~\bibnamefont{Zelnikov}}, \bibnamefont{and}
  \bibinfo{author}{\bibfnamefont{O.}~\bibnamefont{Heinrich}},
  \bibinfo{journal}{Phys. Lett. B} \textbf{\bibinfo{volume}{224}},
  \bibinfo{pages}{255} (\bibinfo{year}{1989}).

\bibitem[{\citenamefont{Frolov et~al.}(1996)\citenamefont{Frolov, Hendy, and
  Larsen}}]{Frolov:1995vp}
\bibinfo{author}{\bibfnamefont{V.~P.} \bibnamefont{Frolov}},
  \bibinfo{author}{\bibfnamefont{S.}~\bibnamefont{Hendy}}, \bibnamefont{and}
  \bibinfo{author}{\bibfnamefont{A.~L.} \bibnamefont{Larsen}},
  \bibinfo{journal}{Phys. Rev. D} \textbf{\bibinfo{volume}{54}},
  \bibinfo{pages}{5093} (\bibinfo{year}{1996}), \eprint{hep-th/9510231}.

\bibitem[{\citenamefont{Igata et~al.}(2018)\citenamefont{Igata, Ishihara,
  Tsuchiya, and Yoo}}]{Igata:2018kry}
\bibinfo{author}{\bibfnamefont{T.}~\bibnamefont{Igata}},
  \bibinfo{author}{\bibfnamefont{H.}~\bibnamefont{Ishihara}},
  \bibinfo{author}{\bibfnamefont{M.}~\bibnamefont{Tsuchiya}}, \bibnamefont{and}
  \bibinfo{author}{\bibfnamefont{C.-M.} \bibnamefont{Yoo}},
  \bibinfo{journal}{Phys. Rev. D} \textbf{\bibinfo{volume}{98}},
  \bibinfo{pages}{064021} (\bibinfo{year}{2018}), \eprint{1806.09837}.

\bibitem[{\citenamefont{Xing et~al.}(2021)\citenamefont{Xing, Levin, Gruzinov,
  and Vilenkin}}]{Xing:2020ecz}
\bibinfo{author}{\bibfnamefont{H.}~\bibnamefont{Xing}},
  \bibinfo{author}{\bibfnamefont{Y.}~\bibnamefont{Levin}},
  \bibinfo{author}{\bibfnamefont{A.}~\bibnamefont{Gruzinov}}, \bibnamefont{and}
  \bibinfo{author}{\bibfnamefont{A.}~\bibnamefont{Vilenkin}},
  \bibinfo{journal}{Phys. Rev. D} \textbf{\bibinfo{volume}{103}},
  \bibinfo{pages}{083019} (\bibinfo{year}{2021}), \eprint{2011.00654}.

\bibitem[{\citenamefont{Cyr et~al.}(2022)\citenamefont{Cyr, Jiao, and
  Brandenberger}}]{Cyr:2022urs}
\bibinfo{author}{\bibfnamefont{B.}~\bibnamefont{Cyr}},
  \bibinfo{author}{\bibfnamefont{H.}~\bibnamefont{Jiao}}, \bibnamefont{and}
  \bibinfo{author}{\bibfnamefont{R.}~\bibnamefont{Brandenberger}}
  (\bibinfo{year}{2022}), \eprint{2202.01799}.

\bibitem[{\citenamefont{Vilenkin et~al.}(2018)\citenamefont{Vilenkin, Levin,
  and Gruzinov}}]{Vilenkin:2018zol}
\bibinfo{author}{\bibfnamefont{A.}~\bibnamefont{Vilenkin}},
  \bibinfo{author}{\bibfnamefont{Y.}~\bibnamefont{Levin}}, \bibnamefont{and}
  \bibinfo{author}{\bibfnamefont{A.}~\bibnamefont{Gruzinov}},
  \bibinfo{journal}{JCAP} \textbf{\bibinfo{volume}{11}}, \bibinfo{pages}{008}
  (\bibinfo{year}{2018}), \eprint{1808.00670}.

\bibitem[{\citenamefont{Brandt and Seidel}(1996)}]{Brandt:1996si}
\bibinfo{author}{\bibfnamefont{S.~R.} \bibnamefont{Brandt}} \bibnamefont{and}
  \bibinfo{author}{\bibfnamefont{E.}~\bibnamefont{Seidel}},
  \bibinfo{journal}{Phys. Rev. D} \textbf{\bibinfo{volume}{54}},
  \bibinfo{pages}{1403} (\bibinfo{year}{1996}), \eprint{gr-qc/9601010}.

\bibitem[{\citenamefont{Kibble and Turok}(1982)}]{Kibble:1982cb}
\bibinfo{author}{\bibfnamefont{T.~W.~B.} \bibnamefont{Kibble}}
  \bibnamefont{and} \bibinfo{author}{\bibfnamefont{N.}~\bibnamefont{Turok}},
  \bibinfo{journal}{Phys. Lett. B} \textbf{\bibinfo{volume}{116}},
  \bibinfo{pages}{141} (\bibinfo{year}{1982}).

\bibitem[{\citenamefont{Burden}(1985)}]{Burden:1985md}
\bibinfo{author}{\bibfnamefont{C.~J.} \bibnamefont{Burden}},
  \bibinfo{journal}{Phys. Lett. B} \textbf{\bibinfo{volume}{164}},
  \bibinfo{pages}{277} (\bibinfo{year}{1985}).

\bibitem[{\citenamefont{Gage}(1984)}]{gage1984curve}
\bibinfo{author}{\bibfnamefont{M.~E.} \bibnamefont{Gage}},
  \bibinfo{journal}{Inventiones mathematicae} \textbf{\bibinfo{volume}{76}},
  \bibinfo{pages}{357} (\bibinfo{year}{1984}).

\bibitem[{\citenamefont{Gage}(1986)}]{gage1986shrinking}
\bibinfo{author}{\bibfnamefont{M.}~\bibnamefont{Gage}}, \bibinfo{journal}{J.
  Differential Geom.} \textbf{\bibinfo{volume}{23}}, \bibinfo{pages}{69}
  (\bibinfo{year}{1986}).

\bibitem[{\citenamefont{Grayson}(1989)}]{grayson1989shortening}
\bibinfo{author}{\bibfnamefont{M.~A.} \bibnamefont{Grayson}},
  \bibinfo{journal}{Annals of Mathematics} \textbf{\bibinfo{volume}{129}},
  \bibinfo{pages}{71} (\bibinfo{year}{1989}).

\end{thebibliography}
\end{document}